\documentclass{aa}  
\usepackage{graphicx}
\usepackage{caption}
\usepackage{color}
\usepackage[english]{babel}
\usepackage{graphicx}
\usepackage{lscape}
\usepackage{multirow}
\usepackage{natbib}
\usepackage[scaled]{helvet}
\usepackage[varg]{txfonts}
\usepackage{url}
\usepackage{xspace}
\usepackage{subfig} 
\usepackage{amssymb}
\usepackage[version=4]{mhchem}
\usepackage{scrextend}
\usepackage{stmaryrd}
\bibpunct{(}{)}{;}{a}{}{,}
\newcommand{\hd}{\object{HD\,49798}\xspace}
\newcommand{\rx}{\object{RX\,J0648.0$-$4418}\xspace}
\newcounter{Rco}
\newcommand{\ionw}[3]{\mbox{\ion{#1}{#2}~$\lambda\,#3\,\mathrm{\AA}$}\xspace}
\newcommand{\ionww}[3]{\mbox{\ion{#1}{#2}~$\lambda\lambda\,#3\,\mathrm{\AA}$}\xspace}

\newcommand{\Jonww}[3]{\mbox{\ion{#1}{#2}~$\lambda\lambda\,#3$\,\AA}\xspace}

\newcommand{\logg}{\mbox{$\log g$}\xspace}
\newcommand{\loggw}[1]{\mbox{$\log g\hspace{-0.5mm} =\hspace{-0.5mm}  #1$}}

\newcommand{\sga}{\raisebox{-0.10em}{$\stackrel{>}{{\mbox{\tiny $\sim$}}}$}}

\newcommand{\sla}{\raisebox{-0.10em}{$\stackrel{<}{{\mbox{\tiny $\sim$}}}$}}
\newcommand{\spm}{\mbox{\raisebox{0.20em}{{\tiny \hspace{0.2mm}\mbox{$\pm$}\hspace{0.2mm}}}}}

\newcommand{\Teff}{\mbox{$T_\mathrm{eff}$}\xspace}
\newcommand{\Teffw}[1]{\mbox{$\Teff\hspace{-0.5mm} =\hspace{-0.5mm} #1 \,\mathrm{K}$}}
\newcommand{\ebv}{\mbox{$E_\mathrm{B-V}$}}

\begin{document}
          \title{Spectral analysis of \hd, \\
                 a bright, hydrogen-deficient sdO-type donor star \\ 
                 in an accreting X-ray binary
                }                

          \titlerunning{Spectral analysis of \hd, a H-deficient sdO-type donor star in an accreting X-ray binary}

   \author{T\@. Rauch
           \and
           P\@. Strau\ss}

\institute{Institute for Astronomy and Astrophysics,
           Kepler Center for Astro and Particle Physics,\\
           Eberhard Karls University,
           Sand 1,
           72076 T\"ubingen,
           Germany,\\
           \email{rauch@astro.uni-tuebingen.de}
           }

   \date{Received 16 April 2025; accepted 25 June 2025}

  \abstract
   {\hd is a bright ($m_{\mathrm{V}} = 8.287$), hot (effective temperature \Teffw{45\,000}) 
    subdwarf star of the spectral type O (sdO).
    It is the only confirmed sdO-type mass-donor star of an X-ray binary that has a high-mass ($1.28\,M_\odot$)
    white-dwarf or neutron-star primary with a spin period of only 13.2\,s.}
   {Since a high-quality spectrum of \hd, obtained with the T\"ubingen Ultraviolet Echelle Spectrometer (TUES),
     that has never been analyzed before is available in the database of the Orbiting and Retrievable Far and Extreme Ultraviolet Spectrometer (ORFEUS), we performed a spectral analysis based on observations from
    the far ultraviolet (FUV) to the optical wavelength range.  
   }
   {We used advanced non-local thermodynamic equilibrium (NLTE) model atmospheres of the T\"ubingen Model-Atmosphere
    Package (TMAP) to determine the 
    effective temperature, the surface gravity (\logg), and the
    abundances of those elements that exhibit lines in the available observed spectra.
    }
   {We determined 
    \Teffw{45\,000 \pm 1\,000},
    \loggw{4.46 \pm 0.10},
    and re-analyzed the previously determined photospheric abundances of 
    H, He, N, O, Mg, Al, Si, Fe, and Ni.
    For the first time, we measured the abundances of C, Ne, P, S, Cr, and Mn.
    }
   {Our panchromatic spectral analysis of \hd -- from the FUV to the optical -- allowed us to reduce the error limits
    of the photospheric parameters and to precisely measure the metal abundances. \hd is a stripped, intermediate-mass 
    (zero-age main sequence mass of $M_\mathrm{ZAMS} \approx 7.15\,M_\odot$) He star with a mass of 
    $1.14^{+0.30}_{-0.24}\,M_\odot$. At its surface, it
    exhibits abundances that are the result of CNO-cycle and 3\,$\alpha$ burning nucleosynthesis as well as 
    enhanced Cr, Mn, Fe, and Ni abundances.
    }


   \maketitle
%

\section{Introduction}
\label{sect:intro}

\hd is located in the constellation Puppis and 
was first mentioned in the Henry Draper Catalogue \citep{cannonetal1918}
as an Oe5-type with Ptm and Ptg brightnesses of 8.6 and 7.6, respectively.
It exhibits a very strong $\zeta$ Puppis series of \ion{He}{ii} lines.
It was then observed by \citet{oosterhoff1951}, who found a photo-electric magnitude of $m=8.23$
in the yellow spectral region (filter 3385). \citet{landoltuomoto2007} measured brightnesses for
stars used as spectrophotometric standards for the \textit{Hubble} Space Telescope (HST) and found
$m_\mathrm{V}=8.287$ for \hd (which was too bright for their original observing plan).

\citet{feastetal1957} identified a very strong \ion{He}{ii} Pickering series and a variable velocity.
\citet{jascheketal1963} analyzed optical spectra and reported then on the discovery of a 
new O-type subdwarf (sdO), which was the brightest known at that time and -- even more important -- 
in a spectroscopic binary.

\begin{figure}[t]
        \begin{center}
                \includegraphics[width=\columnwidth]{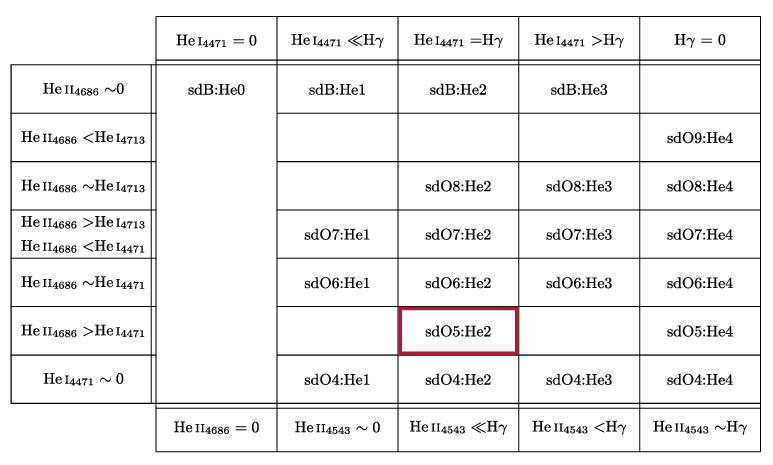}
                \caption{Classification scheme for hot subdwarf stars by \citet{jefferyetal1997}. 
                The red box denotes the classification of \hd.}
                \label{fig:classes}
        \end{center}
\end{figure}

By comparing the spectra of \ion{H}{i}, \ion{He}{i}, and \ion{He}{ii} dominated helium-rich hot subdwarfs, 
\citet{jefferyetal1997} proposed a classification scheme (Fig.\,\ref{fig:classes}) to further differentiate 
between different types of sdOs. 

\hd is difficult to place in this scheme, as it deviates from the typical spectral features described by \citet{heber2016}. 
\ionw{He}{i}{4471} and \ionw{He}{ii}{4543} are slightly shallower than  H\,$\gamma$. This makes the sdOX:He2 classification 
the only reasonable one although the features of the spectrum are not described perfectly. 
The line-depth relation of \ionw{He}{ii}{4686} > \ionw{He}{i}{4471} gives a final classification of sdO5:He2.

\begin{figure}
   \resizebox{\hsize}{!}{\includegraphics{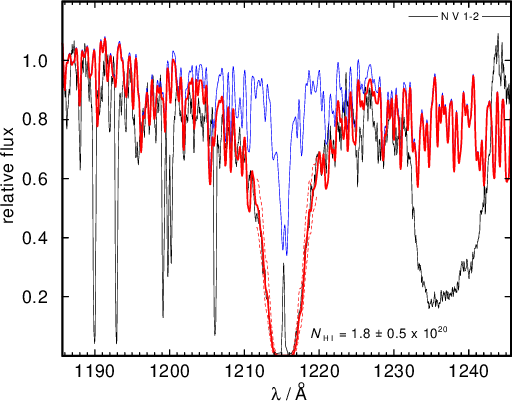}}
           \caption
           {Section of the TUES observation around Ly\,$\alpha$ (black).
            The interstellar \ion{H}{i} absorption was calculated with
            $N_{H\,I}$ = 1.8 (full, red) $\pm 0.5$ (dashed, red, 2.5 times the estimated error range for clarity) 
            $\times 10^{20} \mathrm{cm}^{-2}$.
            The pure stellar spectrum is shown in blue.
            A rotational line broadening ($v_\mathrm{rot} = 37.5\,\mathrm{km/s}$) is applied.
            Interstellar metal lines are not modeled in this plot. 
            The P Cygni profile of the \ion{N}{v} resonance line is marked at the top.
            }
   \label{fig:nhi}
\end{figure}

\begin{figure}
   \resizebox{\hsize}{!}{\includegraphics{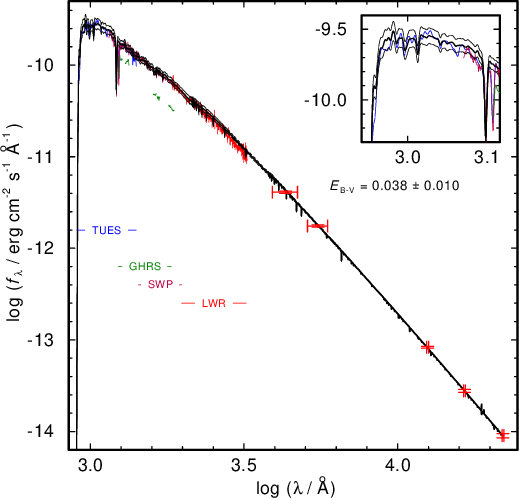}}
           \caption
           {TUES, GHRS, and IUE observations compared with our final model
            normalized to the 2MASS H magnitude.
            \ebv = $0.038\pm 0.010$ is applied.
            In addition, the B and V magnitudes \citep{landoltuomoto2007}
            as well as the 2MASS J, H, and K$_\mathrm{s}$ magnitudes \citep{cutrietal2003} are shown.
            For clarity, TUES and GHRS spectra are convolved with a Gaussian with a full width at half maximum (FWHM) of 7\,\AA.}
   \label{fig:ebv}
\end{figure}

A first quantitative analysis of \hd was presented by \citet{richter1971}, who used a
local thermodynamical equilibrium (LTE) approach with flux-constant, unblanketed models to
fit the observed lines. He found an effective temperature of \Teffw{45\,000}, a surface
gravity of $\log (g\,/\,\mathrm{cm/s^2}) = 5$, $\log \epsilon_\mathrm{He} = 11.8$, and $\log \epsilon_\mathrm{H} = 9.7$
(relative to $\log \epsilon_\odot = 12$, i.e., a number ratio $N_\mathrm{He}/N_\mathrm{H} = 0.63$). 
He estimated a stellar mass of $M = 2\,M_\odot$.
\citet{dufton1972} used optical spectra and investigated non-LTE (NLTE) effects. He arrived at
\Teffw{45\,000 \spm 5\,000}, \loggw{4.3 \spm 0.3}, and a He overabundance of five.
\citet{kudritzki1976} calculated a grid of H+He composed NLTE model atmospheres and derived
\Teffw{43\,700}, \loggw{4.1} and \Teffw{38\,000}, \loggw{4.3} for the two grid abundances
$N_\mathrm{He}/N_\mathrm{H} = 0.1$ and $N_\mathrm{He}/N_\mathrm{H} = 1.0$, respectively.
This was refined in an individual complete NLTE analysis by \citet{kudritzkisimon1978} with H+He models.
They measured 
\Teffw{47\,500 \spm 2\,000}, \loggw{4.25 \spm 0.2} and $N_\mathrm{He}\,/\,N_\mathrm{H} = 0.5 \spm ^{10}_{~7}\mathrm{\%}$.

From the P\,Cygni line profile of the \ion{N}{v} resonance line (cf\@{.} Fig.\,\ref{fig:nhi})
in a spectrum obtained with the International Ultraviolet
Explorer (IUE), \citet{hamannetal1981} determined a
mass-loss rate of $-9.3 \sla (\dot{M}\,/\,M_\odot/\mathrm{yr}) \sla 8.0$ and a terminal wind velocity of
$v_\infty = 1\,350\,\mathrm{km/s}$.

\hd is the sdO donor in a binary system \citep[orbital period $P \approx 1.55\,\mathrm{d}$,][]{thackeray1970}
for a pulsating X-ray source (\rx, $M \approx 1.2\,M_\odot$). It
may be either a massive white dwarf (WD) or a neutron star \citep{mereghettietal2009}.

To constrain the previous and future evolution of this system and to determine the nature of \hd 's companion, 
\citet{brooksetal2017} performed various  
MESA\footnote{Modules for Experiments in Stellar Astrophysics, \citet{paxtonetal2011, paxtonetal2013, paxtonetal2015}} simulations. 
They used \Teffw{47\,500} of \citet{kudritzkisimon1978} for their calculations.
Two scenarios for the origin of sdOs in binaries are most likely.
One, in which the companion is
a neutron star, is a core-collapse supernova of a star with a 
zero-age main sequence (ZAMS) mass $\sga 10\,M_{\odot}$.
A detailed description can be found in \citet{branchwheeler2017}, \citet{nomoto1987}, e.g., stars with 
ZAMS masses of $8 - 10 M_{\odot}$ 
undergo He and C burning, which leaves behind a non-degenerate O + Ne + Mg core, light enough that no Ne burning occurs. 
The subsequent evolution consists of a H-He double-shell burning. As soon as the core (including the C + O layer) reaches 
the critical mass of $1.375\,M_{\odot}$, electron capture on \ce{^{20}F}, \ce{^{20}Ne}, \ce{^{24}Na}, and \ce{^{24}Mg} 
takes place that induces a rapid contraction of the core. The outward traveling shock due to the bounce-back of the shell on 
to the core gives rise to the supernova explosion, leaving behind a neutron star.

An sdO scenario that also fits the observations consists of a $7.15\,M_{\odot}$ ZAMS star that enters a common 
envelope (CE) with its compact companion just before the second dredge-up on the early asymptotic giant branch. 
These simulations hold for both types of compact companions as the evolution of \hd in this scenario is the same up until 
the Roche-lobe overflow (RLOF) and is similar afterwards.

The X-ray spectrum and the nature of the compact companion star were the focus of several
studies \citep{mereghettietal2009,mereghettietal2011,mereghettietal2013,mereghettietal2021,rigosellietal2023} and
strongly improved our evolutionary picture. \citet{brooksetal2017} investigated its past binary
interaction and estimated a time of 65\,000\,yrs from now for the \hd to fill its Roche lobe. Then
the companion will accrete at a rate of about $10^{-5} M_\odot/\mathrm{yr}$. 
Recently, \citet{krtickaetal2019} analyzed new optical spectra obtained with the
Ultraviolet and Visual Echelle Spectrograph (UVES) of the European Southern Observatory (ESO).
In addition, they used spectra taken with the Goddard High-Resolution Spectrograph (GHRS) that
had been aboard the HST. They derived
\Teffw{45\,900 \spm 800}, $M = 1.46 \spm 0.32\,M_\odot$, and determined the photospheric
abundances of H, He, N, O, Mg, Al, Si, Fe, and Ni. Since the 
Orbiting and Retrievable Far and Extreme UV Spectrometer (ORFEUS) spectrum of \hd,
obtained with the T\"ubingen Ultraviolet Echelle Spectrometer (TUES),
had been neglected, we decided to perform a detailed spectral analysis with advanced
NLTE model-atmosphere techniques.

In Sect.\,\ref{sect:obs}, we briefly describe the observations, followed by an overview of the atomic
data used and our NLTE model-atmosphere code in Sect.\,\ref{sect:models}. The analysis is presented in Sect.\,\ref{sect:analysis}.
We summarize our results and conclude in Sect.\,\ref{sect:results}.

\begin{figure}
   \resizebox{\hsize}{!}{\includegraphics{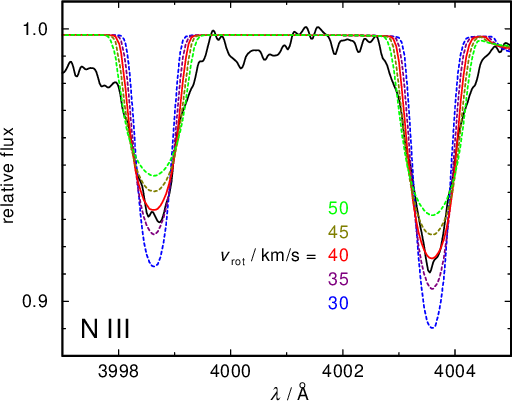}}
           \caption
           {Our final model spectrum with different $v_\mathrm{rot}$ 
            (30 - 50\,km/s, the labels' colors refer to the respective spectra) applied,
            compared with a section of the UVES observation around \ionww{N}{iii}{3998.63,4003.58}.}
   \label{fig:vrot}
\end{figure}

\section{Observations, interstellar reddening, and line absorption}
\label{sect:obs}

ORFEUS\footnote{Orbiting and Retrievable Far- and Extreme Ultraviolet Spectrometer} was an 
ultraviolet (UV) telescope with an 1\,m mirror and a high resolving power $R = \lambda/\Delta\lambda \approx 10\,000$. 
On its second flight (Nov 19 -- Dec 7, 1996, the so-called \mbox{ORFEUS\,II} mission), \hd was observed for 1292\,s
(ObsId 2255\_1, start time Dec 3 1996, 08:04:36 GMT) with the T\"ubingen Echelle Spectrograph (TUES). 
The count rate within $904.63 \le \lambda/\AA \le 1410.45$ was 
6221.3\,/\,s. We used the spectrum (orders 40 to 61) from our institute's 
database\footnote{\url{https://uni-tuebingen.de/fileadmin/Uni_Tuebingen/Fakultaeten/MathePhysik/Institute/IAAT/AIT/Projekte/ORFEUS/Daten/list_of_observations.html}}. 
The same data is available
at the Barbara A\@. Mikulski Archive for Space Telescopes 
(MAST)\footnote{\url{https://archive.stsci.edu/missions-and-data/orfeus/tues}}.

In addition, we used the same UV observation taken with GHRS
aboard the HST and the
optical spectrum ($3752 \le \lambda/\AA \le 5000$) like \citet{krtickaetal2019}. 
The latter was obtained with the UVES at the Very Large Telescope (VLT) of
the ESO on Oct 2, 2016. 
The exposure time was 180\,s and a $0\farcs6$ slit was used. With a resolution-slit product of 41\,400,
we assume a resolving power of $R = \lambda/\Delta\lambda \approx 69\,000$. 

A summary of the observed spectra of \hd used in this analysis is given in Table\,\ref{tab:spectra}.
A comparison of the observations shows that the GHRS spectra are significantly lower in their flux level
(Fig.\,\ref{fig:ebv}): the two spectra with $\lambda < 1400\,\mathrm{\AA}$ (Table\,\ref{tab:spectra}) by a factor of about 1.4, the
others by a factor of about 1.8. Most probably, the pipeline flux-calibration procedure went wrong for these
six GHRS observations.

\begin{table*}
\centering
\caption{Observed spectra used in this analysis.}
\label{tab:spectra}
\begin{tabular}{llr@{\,-\,}lcrl}                                 
\hline
\hline
\noalign{\smallskip}
Instrument                  & Observation Id                        & \multicolumn{2}{c}{Wavelength interval\,/\,\AA} & $R$      & $v_\mathrm{rad}$ & comment                  \\
\hline                                                                                                                            
 \sc{IUE}                   & swp08277                              &  1150 & 1979                                    & 7\,\AA   &                 & Fig.\,\ref{fig:ebv} only \\
                            & lwr07214                              &  1851 & 3349                                    & 7\,\AA   &                 & Fig.\,\ref{fig:ebv} only \\
 \sc{TUES}                  & \sc{tues}2255\_1                      &  \hbox{}\hspace{10mm} 904 & 1410                & 10\,000  &           137.5 & \\
 \textsc{HST/GHRS}          & z0x60606t                             &  1248 & 1270                                    & 25\,000  &           137.5 & \\         
                            & z0x60607t                             &  1300 & 1335                                    & 25\,000  &           137.5 & \\
                            & z0x60608t                             &  1602 & 1637                                    & 25\,000  &           137.5 & \\
                            & z0x60609t                             &  1650 & 1684                                    & 25\,000  &           137.5 & \\
                            & z0x6060at                             &  1800 & 1834                                    & 25\,000  &           137.5 & \\
                            & z0x6060ct                             &  1840 & 1874                                    & 25\,000  &           137.5 & \\
 \textsc{UVES} at \textsc{UT2} & \textsc{ADP}.2020-08-14T15:14:31.993  &  3050 & 3868                                    & 40\,970  &             6.0 & \\                    
 \textsc{UVES} at \textsc{UT2} & \textsc{UVES}.2016-10-02T09:03:11.546 &  3732 & 5000                                    & 69\,000  &             6.0 & \citet{krtickaetal2019}  \\
\hline
\end{tabular}
\tablefoot{
The respective wavelength ranges as well as their spectral resolution (for IUE low-resolution
spectra) and resolving power $R=\lambda/\Delta\lambda$ are given in columns 3 and 4, respectively.
}
\end{table*}

In all of our plots, the observations are shifted to rest wavelength.
To slightly smooth the UV observations, they are processed with a low-pass filter \citep{savitzkygolay1964}.

\begin{figure}
   \resizebox{\hsize}{!}{\includegraphics{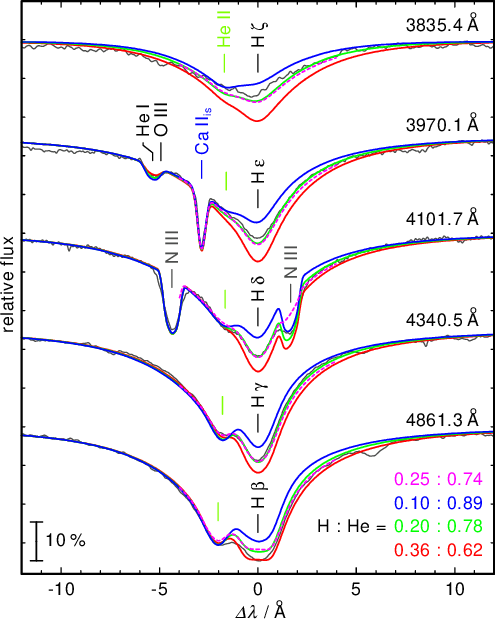}}
           \caption
           {Sections of the observed optical spectrum around \ion{H}{i} Balmer lines and 
            their respective \ion{He}{ii} blends compared with three synthetic spectra with 
            \Teffw{45\,000} and different
            H and He abundances (given in mass fraction, the labels' colors refer to the respective spectra).
            A synthetic spectrum calculated with the photospheric parameters of \citet{krtickaetal2019}
            is shown in the line centers with a dashed, violet line.
            Prominent stellar and interstellar (subscript is) lines are marked.}
   \label{fig:HHe_abund}
\end{figure}

\begin{figure}
   \resizebox{\hsize}{!}{\includegraphics{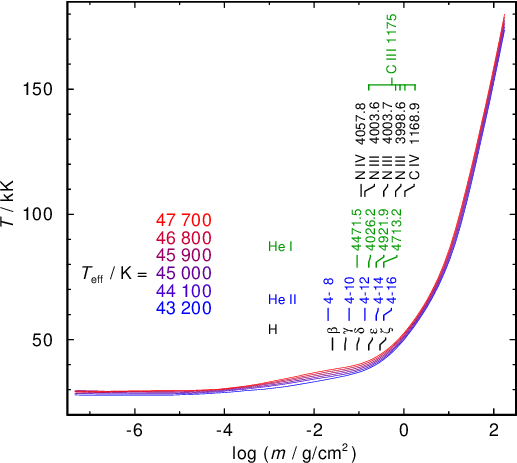}}
           \caption
           {Temperature structures of our models with \loggw{4.46} and different \Teff. 
            The formation depths of the line cores of selected 
            \ion{H}{i}, 
            \ion{He}{i}, \ion{He}{ii}.
            \ion{C}{iii}, \ion{C}{iv},
            \ion{N}{iii}, and \ion{N}{iv}
            lines
            (visible in our TUES or UVES observation) in our \Teffw{47\,000} model are
            indicated.
            $m$ is the column mass, measured from the outer boundary of our model atmospheres.}
   \label{fig:formation}
\end{figure}

To determine the interstellar reddening, we applied the reddening law of 
\citet{fitzpatrick1999} to our final synthetic stellar spectrum and normalized it to the
2MASS H magnitude. We achieved the best agreement to the UV observations at \ebv = 0.038
within a small error range of 0.010  (Fig.\,\ref{fig:ebv}). 

The interstellar neutral hydrogen column density was measured from a comparison of our
theoretical \ion{H}{i} Lyman\,$\alpha$ line profile with the TUES observation.
We reproduced the observation best at $N_{H\,I} = 1.8 \pm 0.2 \times 10^{20} \mathrm{cm}^{-2}$
(Fig.\,\ref{fig:nhi}). This value is in agreement with a reddening value of
\ebv = $0.038\pm 0.010$
if we consider the empirical relation for the Galactic reddening of \citet[$\log (N_\ion{H}{i}/\ebv) = 21.58\pm 0.10$]{groenewegenlamers1989}
and arrive at \ebv = $0.047^{+0.019}_{-0.014}$.

For the detailed calculation of the interstellar line absorption, we used the line-fitting procedure OWENS
\citep{hebrard02,hebrard03}. It can model different clouds in the interstellar medium (ISM) 
with individual radial and turbulent velocities, temperatures, column densities, and chemical compositions. 
It fits Voigt profiles to the observation using a $\chi^2$ minimization.

\section{Atomic data, model atmospheres, and line broadening}
\label{sect:models}

Although prominent P\,Cygni line profiles of \ion{N}{iv}, \ion{N}{v}, and \ion{S}{vi} are exhibited in the TUES spectrum 
of \hd \citep[a weak stellar wind was measured by][Sect.\,\ref{sect:intro}]{hamannetal1981}, 
most of the lines are obviously not contaminated by wind effects. Thus, we decided to use the
T\"ubingen NLTE model-atmosphere package \citep[TMAP,][]{tmap2012}
to calculate plane-parallel models in radiative and hydrostatic equilibrium. These consider the opacities of
H, He, C, N, O, Ne, Mg, Al, Si, P, S, Ca, Sc, Ti, V, Cr, Mn, Fe, Co, and Ni. Level dissolution (pressure ionization) following
\citet{hummermihalas1988} and \citet{hubenyetal1994} is considered.

For elements with an atomic number $Z$ below 20, the atomic data were compiled from our
T\"ubingen Model Atom Database \citep[TMAD,][]{rauchdeetjen2003}, which 
was constructed as part of the T\"ubingen contribution to the 
German Astrophysical Virtual Observatory (GAVO).
For $Z \ge 20$, we used a statistical approach using super levels and super lines
\citep{rauchdeetjen2003} and calculated the atomic data via our
Iron Opacity and Interface \citep[IrOnIc,][]{muellerringatPhD2013}.

For the detailed line-profile calculations, we used Stark-broadening tables of 
\citet[][extended tables of 2015, priv\@. comm\@.]{tremblaybergeron2009} and \citet{schoeningbutler1989}
for the \ion{H}{i} and \ion{He}{ii} lines, respectively.
Stark line-broadening tables from 
\citet{bcs1974_4471,bcs1975_4921} and were used for \ionww{He}{i}{4471, 4921}, respectively,
\citet{griem_1974} for other \ion{He}{i} lines, and
\citet{dimitrijevic1992_CIV,dimitrijevic1992_NV,dimitrijevic1995_OIV-V,dimitrijevic1991_CIV_SIV} 
were used for \ion{C}{iv}, \ion{N}{v}, \ion{O}{iv} - \ion{}{v}, and \ion{Si}{iv} lines, respectively.

The sharp metal lines that are broadened due to the quadratic Stark effect show the impact of rotational
broadening in their line cores. As an example, we show the prominent \ionww{N}{iii}{3998.63, 4003.58} lines
of the 4d\,$^\mathrm{2}$D -- 5f\,$^\mathrm{2}$F$^\mathrm{o}$ multiplet in Fig.\,\ref{fig:vrot}. 
The average determined projected rotational velocity from these and many other lines 
is $v_\mathrm{rot} = 37.5 \pm 5\,\mathrm{km/s}$. 
This matches the value  $v_\mathrm{rot} = 40 \pm 5\,\mathrm{km/s}$ measured by \citet{krtickaetal2019}.

\begin{figure}
   \resizebox{\hsize}{!}{\includegraphics{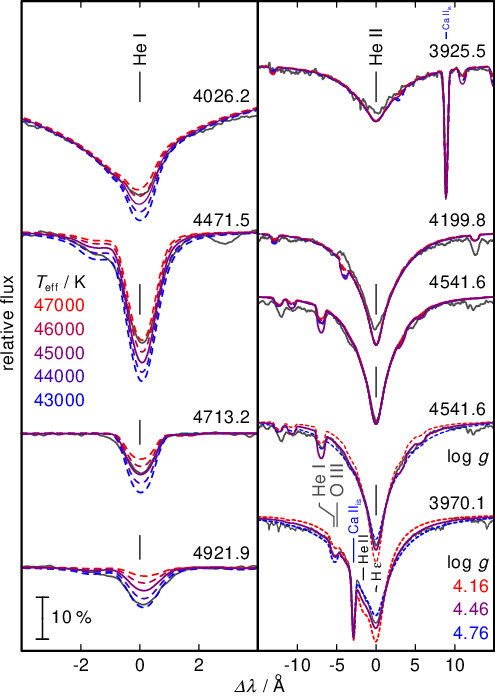}}
           \caption
           {Sections of the observed optical spectrum around prominent \ion{He}{i} (\emph{left panel})
            and \ion{He}{ii} (\emph{right}) lines
            compared to synthetic spectra with different \Teff (the labels' colors refer
            to the respective spectra). The two lower \ion{He}{ii} lines are compared with synthetic
            spectra with different \logg.}
   \label{fig:HHe_Tefflogg}
\end{figure}

\section{Analysis}
\label{sect:analysis}

For our first test models, we adopted the parameters of 
\citet[\Teffw{45\,900}, 
       \loggw{4.56}]{krtickaetal2019}, and used the elements (H, He, N, O, Mg, Al, Si, Fe, and Ni)
with detailed abundance analysis of \citet{krtickaetal2019}. For C, we used their upper limit.
Other elements that were mentioned by \citet{krtickaetal2019} and included in their models with solar abundance values were
neglected here. We used their observed optical spectrum (Fig.\,\ref{fig:HHe_abund}) for comparison. 

\citet{krtickaetal2019} determined H and He mass fractions of 0.25 and 0.74, respectively, with an uncertainty 
of $\pm 10$\,\%. At this H/He abundance ratio, all \ion{H}{i} Balmer-series line cores but that of H\,$\beta$ 
are well reproduced with our test model (Fig.\,\ref{fig:HHe_abund}, with dashed violet lines). 
With our TMAP models (\Teffw{45\,000}, \loggw{4.46}, see their determinations below), we reproduced the 
individual line cores of H\,$\gamma$ - $\zeta$ well at H and He abundances of 0.20 and 0.78 $\pm 2$, respectively 
(Fig.\,\ref{fig:HHe_abund}). This is, within error limits, in agreement with the result of \citet{krtickaetal2019}.

The too-shallow line core of H\,$\beta$ (Fig.\,\ref{fig:HHe_abund}) in our model is unexplained. 
A significant impact of the so-called Balmer-line problem \citep[e.g.,]{napiwotzkirauch1994} 
can be excluded, since we considered many metals in detail (Sect.\,\ref{sect:models}). 
The Balmer-line problem arises if a too-low metal opacity is considered in the models,
e.g., in pure H+He models. In such models. the resulting too-high temperature in the outer atmospheres where 
the lower members of the Balmer series form (Fig.\,\ref{fig:formation}) makes it impossible to 
reproduce all observed Balmer-line profiles with the same \Teff. 

We adopted our H/He abundance values of 0.20/0.79 from the H\,$\gamma$ - $\zeta$ / \ion{He}{ii} blends. We checked 
the abundances during our further analysis after any change in the atmospheric parameters. 

\paragraph{The effective temperature} was determined from ionization equilibria, i.e.,
by the evaluation of the equivalent-width ratios of lines of at least two subsequent
ionization stages -- this is a very sensitive indicator for \Teff.
Thus, we calculated models with different \Teff (43\,000\,K - 49\,000\,K, Fig.\,\ref{fig:HHe_Tefflogg}). 
We started with the \ion{He}{i}/\ion{He}{ii} ionization equilibrium and verified this result
by other ionization equilibria of metals, where we could even identify lines of three subsequent
ionization stages.

While the \ion{He}{ii} line profiles do not change significantly in our above \Teff interval, the \ion{He}{ii} lines are 
dependent on \Teff. We do not have a perfect agreement at one \Teff, but $45\,000\pm 1\,000$\,K was a good
start value for the analysis of the ORFEUS\,II observation. This is in agreement with the result of 
\citet[][\Teffw{45\,900 \spm 800}]{krtickaetal2019}.
Other ionization equilibria were checked in the flow of our analysis to verify this preliminary result and to reduce
the error range.

\begin{figure}
   \resizebox{\hsize}{!}{\includegraphics{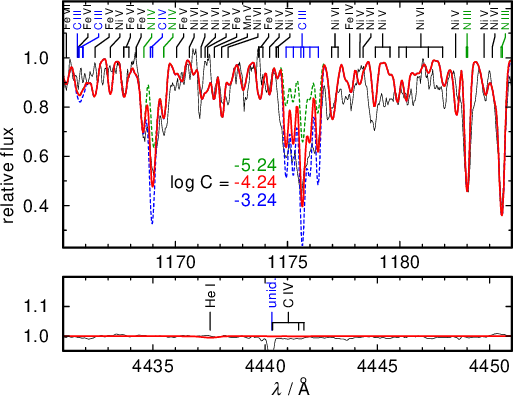}}
           \caption
           {\emph{Top panel:} Section of the TUES observation around 
            \ionww{C}{iii}{1174.93 - 1176.37} and \ionww{C}{iv}{1168.89, 1168.98}
            compared with our synthetic spectra with different C mass fractions
            (labeled with the logarithm, their colors indicate the respective synthetic spectra).
            \emph{Bottom:} Same as top panel, for a section of the UVES observation around 
            \ionw{He}{i}{4437.55} and \ionww{C}{iv}{4440.33 - 4441.73}.
            }
   \label{fig:carbon}
\end{figure}

\paragraph{The surface gravity} was examined using \ion{H}{i} and \ion{He}{ii} lines within $4.16 \sla \log g \sla 4.76$
\citep[starting with the value of \loggw{4.56\pm 0.08} of][]{krtickaetal2019}. An example of our comparison is
shown in Fig.\,\ref{fig:HHe_Tefflogg}. While the line cores are too deep and the line wings are too narrow at the lower values,
the central depression of the lines in the observation is not reproduced at the higher values. We adopted \loggw{4.46\pm 0.1}.
This is well in agreement with \citet{krtickaetal2019}.

With our new parameters, we improved the agreement between the synthetic and observed spectra 
\citep[cf\@. Fig.\,1 of][]{krtickaetal2019}. From our new models, we calculated spectra for the
analysis of the TUES, GHRS, and UVES observations.

\begin{figure}
   \resizebox{\hsize}{!}{\includegraphics{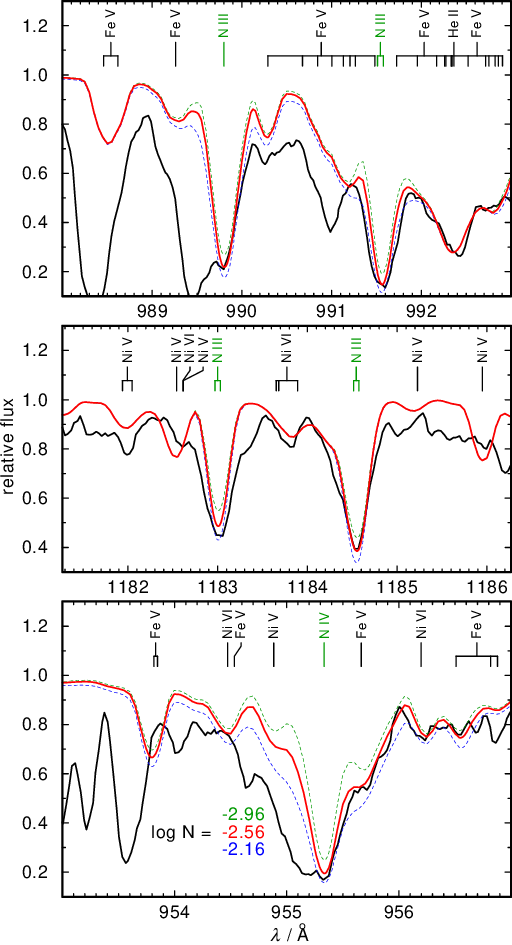}}
           \caption
           {Same as Fig.\,\ref{fig:carbon} but with the TUES observation and different N mass fractions
            around 
            \ionww{N}{iii}{989.80, 991.51, 991.58} (\emph{top panel}),           
            \ionww{N}{iii}{1182.97, 1183.03, 1184.51, 1184.57} (\emph{middle}), and
            \ionw{N}{iv}{954.47} (\emph{bottom}).
            }
   \label{fig:nitrogen}
\end{figure}

\begin{figure}
   \resizebox{\hsize}{!}{\includegraphics{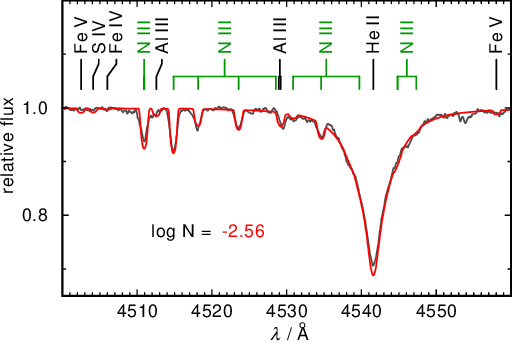}}
           \caption
           {Same as Fig.\,\ref{fig:nitrogen} but showing the UVES observation around 
            \ionww{N}{iii}{4510-4547}.
            }
   \label{fig:nitrogen_opt}
\end{figure}

\begin{figure}
   \resizebox{\hsize}{!}{\includegraphics{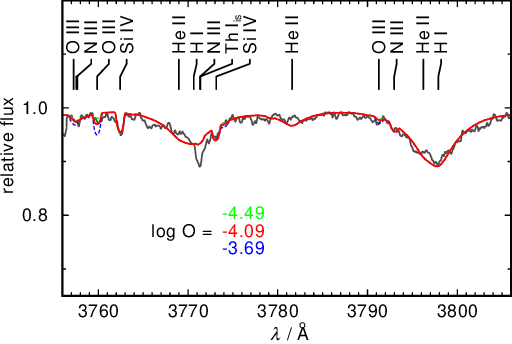}}
           \caption
           {Same as Fig. \ref{fig:nitrogen_opt} but for different O mass fractions
            around
            \ionww{O}{iii}{3757.23, 3759.88, 3791.28}.
            }
   \label{fig:oxygen_opt}
\end{figure}

In the following, we determine the abundances of C, N, O, Ne, Mg, Al, Si, P, S, Cr, Mn, Fe,  and Ni
using this strategy. In the case that lines of an element are exhibited in our model, we try to identify their
observed counterparts. Since the rotational velocity ($v_\mathrm{rot} = 40\,\mathrm{km/s}$) results in a
significant line broadening, wavelengths of lines that contribute to absorption features can be measured from the
unrotated synthetic spectrum. In most figures, 
a thin, light green line shows a synthetic spectrum, which was calculated without the opacities of the
respective element. The comparison to the red line, which is a synthetic spectrum that was calculated from
the same model considering the opacities of that element, allows us to identify its line-opacity 
contributions even in blended line features.

Subsequently, we adjusted the model abundances to reproduce the observation well. 
Table\,\ref{tab:lines} summarizes the strategic lines used in this analysis. With the improved 
abundances, the model was then recalculated to account for the impact of the changed opacities on the atmospheric
structure and verify the metal abundances (as well as \Teff, \logg, and the other abundance ratios). 
To determine the abundance uncertainties due to the error propagation from \Teff and \logg, we measured the
abundances from two models at the error limits of \Teff and \logg, namely the 
\Teff + $\Delta$\Teff / \logg - $\Delta$\logg and the 
\Teff - $\Delta$\Teff / \logg + $\Delta$\logg models (the $\Delta$s are the error ranges),
for the highest- and lowest-ionization models, respectively.

\begin{table*}\centering
\caption{Wavelengths of the strongest metal lines used for the abundance determination.}
\label{tab:lines}
\setlength{\tabcolsep}{0.4em}
\begin{tabular}{rlp{16cm}}                                             
\hline
\hline
\noalign{\smallskip}
\multicolumn{2}{c}{Ion} & Wavelength / \AA          \\
\hline
\noalign{\smallskip}
C  & \sc{iii}           & 1174.93\tablefootmark{b},
                          1175.26\tablefootmark{b},
                          1175.59\tablefootmark{b},
                          1175.71\tablefootmark{b},
                          1175.99\tablefootmark{b},
                          1176.37\tablefootmark{b}  \\
   & \sc{iv}            & 1168.89\tablefootmark{b}, 
                          1168.98\tablefootmark{b}  \\
N  & \sc{iii}           & 
                           979.84\tablefootmark{b},
                           979.92\tablefootmark{b},
                           989.79\tablefootmark{b},
                           991.51\tablefootmark{b},
                           991.59\tablefootmark{b},
                          1182.97\tablefootmark{b},
                          1183.03\tablefootmark{b},
                          1184.51\tablefootmark{b},
                          1184.57\tablefootmark{b},
                          3998.63\tablefootmark{a}, 
                          4003.58\tablefootmark{a}, 
                          4097.33\tablefootmark{a}, 
                          4103.43\tablefootmark{a},
                          4195.76\tablefootmark{b},
                          4200.10\tablefootmark{b},
                          4332.91\tablefootmark{b},
                          4345.68\tablefootmark{b},
                          4379.11\tablefootmark{a}, 
                          4510.91\tablefootmark{a}, 
                          4514.86\tablefootmark{a}, 
                          4518.14\tablefootmark{a}, 
                          4523.56\tablefootmark{a},
                          4534.58\tablefootmark{b}, 
                          4634.14\tablefootmark{b}, 
                          4640.64\tablefootmark{b}, 
                          4858.82\tablefootmark{b},
                          4867.15\tablefootmark{b},
                          4873.60\tablefootmark{b}  \\
N  & \sc{iv}            &  955.33\tablefootmark{b},
                          4057.76\tablefootmark{a},
                          4606.33\tablefootmark{b}  \\ 
O  & \sc{iii}           & 3757.23\tablefootmark{a},
                          3759.88\tablefootmark{b}  \\
Ne & \sc{iii}           & 3328.70\tablefootmark{b},
                          3334.84\tablefootmark{b}  \\
Mg & \sc{ii}            & 4481.32\tablefootmark{a}  \\
Al & \sc{iii}           & 1854.72\tablefootmark{a}, 
                          1862.79\tablefootmark{a}, 
                          4479.97\tablefootmark{b}, 
                          4512.56\tablefootmark{b}, 
                          4528.94\tablefootmark{b}  \\
Si & \sc{iv}            & 1122.49\tablefootmark{b}, 
                          1128.32\tablefootmark{b}, 
                          3762.44\tablefootmark{a}, 
                          3773.15\tablefootmark{a}, 
                          4088.85\tablefootmark{a}, 
                          4212.41\tablefootmark{a}, 
                          4631.24\tablefootmark{b}, 
                          4654.32\tablefootmark{a}  \\ 
P  & \sc{v}             & 1118.55\tablefootmark{b}, 
                          1128.01\tablefootmark{b}  \\
S  & \sc{v}             & 1122.03\tablefootmark{b}, 
                          1128.01\tablefootmark{b}  \\
\hline
\end{tabular} 
\tablefoot{Cr, Mn, Fe, and Ni exhibit many lines and are therefore not included here.
           \tablefoottext{a}{Used by \cite{krtickaetal2019}},
           \tablefoottext{b}{additionally used in our analysis}.
           }
\end{table*}

Here, we provide some examples of our abundance analysis.
A complete comparison of observations and synthetic spectra, including a detailed modeling of
the interstellar line absorption,
is shown in Figs.\,\ref{fig:orfeusall} - \ref{fig:uvesall}.
A comparison to the abundance determinations of \citet{krtickaetal2019} is given in
Table\,\ref{tab:parameters} and Fig.\,\ref{fig:abund}. In the following, we use mass fractions for the determined abundances.

\paragraph{The carbon abundance} was measured using the prominent \ionww{C}{iii}{1174.93 - 1176.37} and 
\ionww{C}{iv}{1168.89, 1168.98}.
Lines in the optical wavelength range are much shallower and, thus give a higher uncertainty range for the determined abundances.
We find a good agreement with the TUES observation at a C mass fraction of 
$5.7^{+3.3}_{-2.1}\times 10^{-5}$ (Fig.\ref{fig:carbon}, for clarity, we show $\pm 1$\,dex). 
In addition, the
ionization equilibrium is well matched with our \Teffw{45\,000}/\loggw{4.46} model (Fig.\ref{fig:carbon}).
In the UVES spectrum (Fig.\ref{fig:carbon}, bottom panel), \ionww{C}{iv}{4440.33 - 4441.74}, which are generally
the strongest C lines in the optical, are not visible.

\paragraph{The nitrogen abundance} cannot be determined precisely from a comparison of the TUES observation and synthetic spectra.
Although \ion{N}{iii} and \ion{N}{iv} lines are prominent in the observation and are sufficiently well reproduced by our models,
their line cores appear to be almost saturated (Fig.\,\ref{fig:nitrogen}). The line wings of \ionw{N}{iv}{954.47} are more 
dependent on the N abundance than those of the \ion{N}{iii} lines. An indicator for the N abundance is the red wing of 
\ionw{N}{iv}{954.47}, which is blended by a strong \ion{Fe}{v} line. The center of gravity of this line is visible in the wing 
and well modeled at an N abundance of $2.8^{+1.6}_{-1.0}\times 10^{-3}$ (Fig.\,\ref{fig:nitrogen}, we show $\pm 0.4$\,dex).
At this abundance, the optical observation is very well reproduced (Fig.\,\ref{fig:nitrogen_opt}).

\paragraph{The oxygen abundance} is determined from the most prominent \ion{O}{iii} lines in the UVES
observation, namely \ionww{O}{iii}{3757.23, 3759.88} (Fig.\,\ref{fig:oxygen_opt}). We measure
an O abundance of $8.0^{+0.8}_{-1.1}\times 10^{-5}$.

\paragraph{The iron-group abundances} (here the abundances of calcium to nickel) have a strong impact on the
atmospheric structure. Since at first glance an inspection of our UV observations revealed a high number of iron and nickel
lines, we started with an abundance analysis of these two elements to include their opacities
in our model-atmosphere calculations. The abundances of Ca, Sc, Ti, V, Mn, Cr, and Co are then, in the first step
of our calculations, scaled to the Fe abundance with the solar abundance pattern to model their contribution to the
background opacity. In the course of this analysis, we use then the newly determined values, where available.

\begin{figure}
   \resizebox{\hsize}{!}{\includegraphics{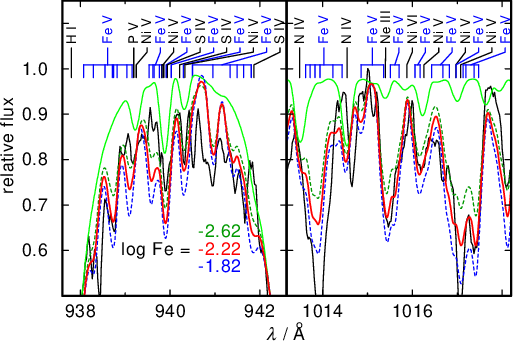}}
           \caption
           {Sections of the TUES observation compared with our synthetic spectra with different Fe abundances
            around some prominent Fe lines. 
            }
   \label{fig:iron}
\end{figure}

\paragraph{The iron abundance} of $6.0^{+3.5}_{-2.2}\times 10^{-3}$
turned out to be significantly higher (about three to seven times in the error range) than the solar abundance 
\citep[$1.3\times 10^{-3}$]{scottetal2015b}.
Since there are so many Fe lines in the UV wavelength region, we show some representatives in
Fig.\,\ref{fig:iron}.

\begin{figure}
   \resizebox{\hsize}{!}{\includegraphics{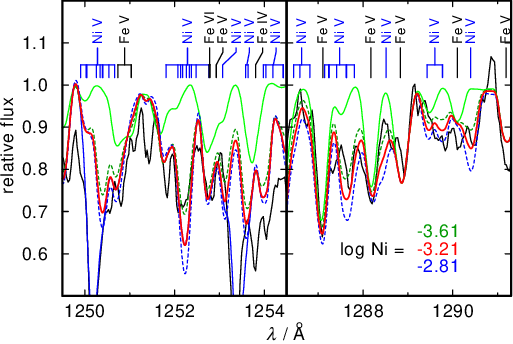}}
           \caption
           {Same as Fig.\,\ref{fig:iron} but with different Ni abundances
            around some prominent Ni lines. 
            }
   \label{fig:nickel}
\end{figure}

\paragraph{The nickel abundance} is also derived from many Ni lines in the UV, we show some examples in
Fig.\,\ref{fig:nickel}. The determined Ni abundance is $6.1^{+2.2}_{-2.2}\times 10^{-4}$.
This is, like found for Fe, also much higher (five to eleven times) than the solar value of $7.6\times 10^{-5}$ \citep{scottetal2015b}.
The Fe/Ni mass-fraction ratio is $9.8^{+14.6}_{-5.2}$, which is lower than the solar ratio of 16.9, but agrees
within its error limits.

\begin{figure}
   \resizebox{\hsize}{!}{\includegraphics{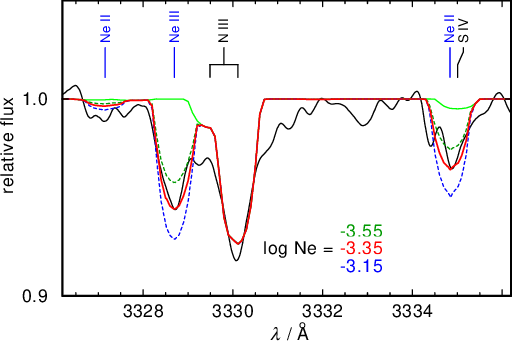}}
           \caption
           {Same as Fig.\,\ref{fig:iron} but with different Ne abundances
            around
            \ionww{Ne}{ii}{3327.153, 3334,837} and \ionw{Ne}{iii}{3328.697}. 
            }
   \label{fig:neon}
\end{figure}

\paragraph{The neon abundance} determination uses \ion{Ne}{ii} and \ion{Ne}{iii} lines (Fig.\,\ref{fig:neon})
that are identified in the UVES spectrum. We find a Ne abundance of $4.5^{+2.5}_{-1.7}\times 10^{-3}$.

\begin{figure}
   \resizebox{\hsize}{!}{\includegraphics{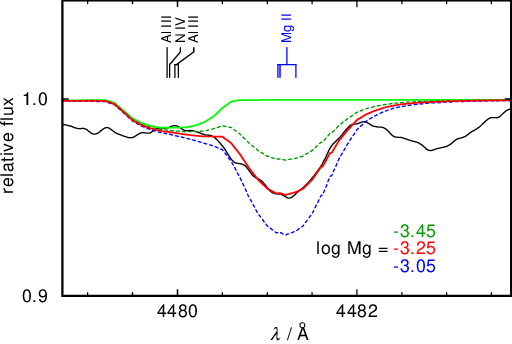}}
           \caption
           {Same as Fig.\,\ref{fig:oxygen_opt} but with different Mg abundances
            around
            \ionww{Mg}{ii}{4481.12, 4481.14, 4481.32}.
            }
   \label{fig:magnesium}
\end{figure}

\paragraph{The magnesium abundance} is measured from the most prominent magnesium line,
namely \ionw{Mg}{ii}{4481.143} (Fig.\,\ref{fig:magnesium}). We arrive at a Mg abundance 
of $5.6^{+2.9}_{-1.8}\times 10^{-4}$.

\begin{figure}
   \resizebox{\hsize}{!}{\includegraphics{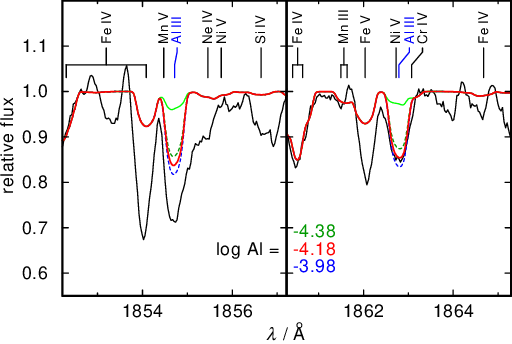}}
           \caption
           {Sections of the GHRS observations compared with our synthetic spectra with different Al abundances
            around
            \ionww{Al}{iii}{1854.72, 1862.79}. 
            }
   \label{fig:aluminum}
\end{figure}

\paragraph{The aluminum abundance} of $5.6^{+2.9}_{-1.8}\times 10^{-4}$ reproduces
well the isolated \ionw{Al}{iii}{1862.79} in the GHRS observation (Fig.\,\ref{fig:aluminum}, right panel). The equally
strong \ionw{Al}{iii}{1854.72} (Fig.\,\ref{fig:aluminum}, left panel) is not suitable for an analysis because it
appears to be blended by interstellar absorption and may have problems during the data-reduction process.

\begin{figure}
   \resizebox{\hsize}{!}{\includegraphics{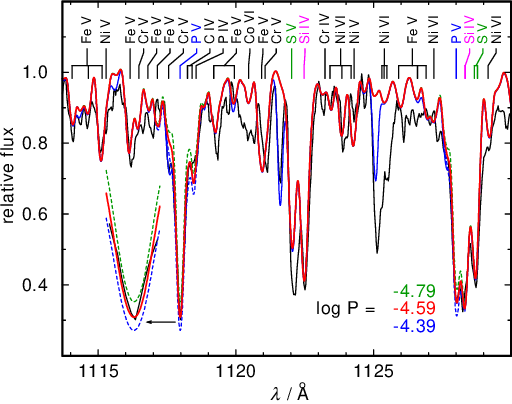}}
           \caption
           {Same as Fig.\,\ref{fig:carbon} but with different P abundances
            around
            \ionww{P}{v}{1117.98, 1128.01} (marked in blue).
            The thin, blue model spectrum includes interstellar line absorption.
            }
   \label{fig:phosphorus}
\end{figure}

\paragraph{The silicon abundance} is $2.1^{+1.2}_{-0.8}\times 10^{-3}$ (Fig.\,\ref{fig:phosphorus}, 
from \ionww{Si}{iv}{1122.49, 1128.33}, marked in magenta).
The TUES observation exhibits many lines of iron-group elements; here we identify
Cr, Fe, and Ni lines.

\paragraph{The phosphorus abundance} is determined from the comparison with the TUES
observation. The prominent \ion{P}{v} lines are best reproduced at
a P abundance of $3.2^{+1.9}_{-1.2}\times 10^{-5}$ (Fig.\,\ref{fig:phosphorus}).

\paragraph{The sulfur abundance} is $2.1^{+1.2}_{-0.8}\times 10^{-3}$ (Fig.\,\ref{fig:phosphorus}, 
from \ionww{S}{v}{1122.03, 1128.67}, marked in green).

\begin{figure}
   \resizebox{\hsize}{!}{\includegraphics{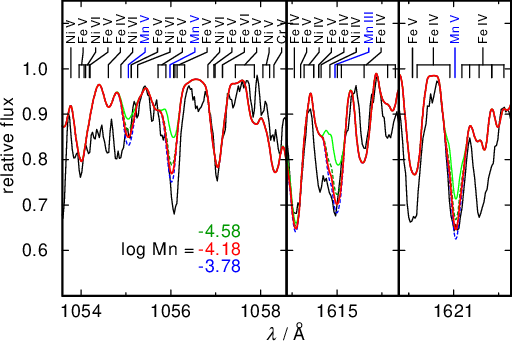}}
           \caption
           {Sections of the TUES  panel) and GHRS (\emph{middle} and \emph{right}) 
            observations compared with our synthetic spectra with different Mn abundances
            around
            \ionww{Mn}{v}{1055.06, 1055.98} (\emph{left}),
            \ionw{Mn}{iii}{1614.96} (\emph{middle}), and
            \ionw{Mn}{v}{1621.03} (\emph{right}). 
            }
   \label{fig:manganese}
\end{figure}

\paragraph{The manganese abundance} can be measured from Mn line contributions to absorption
features in TUES and GHRS observations (Fig.\,\ref{fig:manganese}). Although no isolated Mn
line is identified, we unambiguously identify Mn-line opacities and find
a Mn abundance of $6.6^{+10.0}_{-3.0}\times 10^{-5}$.

\begin{figure}
   \resizebox{\hsize}{!}{\includegraphics{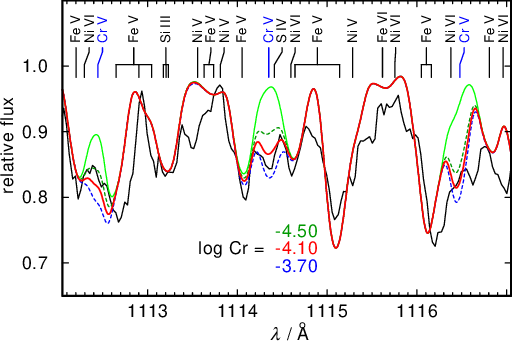}}
           \caption
           {Same as Fig. \ref{fig:phosphorus} but with different Cr mass fractions
            around
            \ionww{Cr}{v}{1112.45, 1114.35, 1116.48}. We show the derived Cr abundance
            $\pm$ 0.4 dex for clarity.
            }
   \label{fig:chromium}
\end{figure}

\paragraph{The chromium abundance} is determined from identified Cr lines in the TUES
observation (e.g., \ionww{Cr}{v}{1116.48}, Fig.\,\ref{fig:chromium}). We find a
Cr abundance of $7.9^{+4.3}_{-3.0}\times 10^{-5}$.

\section{Distance, luminosity, and mass}
\label{sect:dlm}

To determine the distance to a star, one can use the equivalent widths of the interstellar 
H and K lines (\ionww{Ca}{ii}{3968,47, 3933.66}). We measured 
$W_{\lambda}^\mathrm{H} = 122.75\,\mathrm{m\AA}$
$W_{\lambda}^\mathrm{K} = 163.86\,\mathrm{m\AA}$.
These widths can then be used to calculate the distance by using formulae proposed by \citet{megieretal2008},

\begin{align}
        d^\mathrm{H} &= \left( 4.58 \cdot W_{\lambda}^\mathrm{H} +             \mathrm{102} \right) = 664.20\,\mathrm{pc}\quad , \\
        d^\mathrm{K} &= \left( 2.78 \cdot W_{\lambda}^\mathrm{K} + \hspace{1mm}\mathrm{95}  \right) = 550.53\,\mathrm{pc}\quad .
\end{align}

\noindent
The average $d = 607 \pm 80$\,pc is in agreement with the value previously determined by 
\citet[][$650\pm 100$]{kudritzkisimon1978}, 
\citet[][$508\pm 17$]{krtickaetal2019},
and $d = 520.94^{+13.95}_{-13.25}$\,pc, which is calculated from parallax 
measurements\footnote{\textit{Gaia} source\_id: 5562023884304074240} by the \textit{Gaia} mission.

\begin{table}   
        \begin{center}
                \caption[Stellar parameters]{Determined stellar parameters, compared to previously determined values 
                         by \textit{Gaia}, \citet[][K]{krtickaetal2019} and \citet[][KS]{kudritzkisimon1978}.}
                \vspace{1em}
                \setlength{\tabcolsep}{.3em}
                \begin{tabular}{lrlr@{$\, \pm$}lr@{$\, \pm$}lr@{$\, \pm$}l} 
                        \hline
                        \hline
                        &
                        \multicolumn{2}{c}{HD49798}&
                        \multicolumn{2}{c}{\textit{Gaia}}&
                        \multicolumn{2}{c}{K}&
                        \multicolumn{2}{c}{KS}\\
                        \hline
                        \noalign{\smallskip}
                        $d$\,/\,pc     &  496\tablefootmark{a} & $^{+82}_{-83}$     &  521\tablefootmark{b} & 14 &  508\tablefootmark{a} &  17  &  650\tablefootmark{a} &  100 \\
                        \noalign{\smallskip}
                        $L\,/\,L_\odot$ & 4360                  & $^{+509}_{-456}$   & \multicolumn{2}{l}{} & 4400 &  400 & 7943 & 2743 \\
                        \noalign{\smallskip}
                        $M\,/\,M_\odot$ & 1.14                  & $^{+0.30}_{-0.24}$ & \multicolumn{2}{l}{} & 1.46 & 0.32 & 1.75 & 1.0  \\
                        \noalign{\smallskip}
                        $R\,/\,R_\odot$ & 1.09                  & $^{+0.11}_{-0.10}$ & \multicolumn{2}{l}{} & 1.05 & 0.06 & 1.45 & 0.25 \\
                        \noalign{\smallskip}
                        \hline
                \end{tabular}
                \label{tab:para}
                \tablefoot{\tablefoottext{a}\mbox{Spectroscopic distance},
                           \tablefoottext{b}\mbox{trigonometric (parallax) distance}.}
        \end{center}
\end{table}

Alternatively, the distance was calculated utilizing the Eddington flux at $\lambda$ 5454\,\AA $\,$ following the 
formula\footnote{\url{http://astro.uni-tuebingen.de/~rauch/SpectroscopicDistanceDetermination.gif}}

\begin{equation}
d = \mathrm{7.11} \times \mathrm{10}^\mathrm{4}\sqrt{M \cdot H_\nu \cdot \mathrm{10}^{\mathrm{0.4}\cdot m_{{v_0}}-\log g}} \quad \mathrm{pc} \quad ,
\end{equation}

\noindent
where $M$ is the stellar mass in solar units, 
$H_\nu = 6.72^{+0.15}_{-0.14}\times 10^{-4}\,\mathrm{erg/cm^2/s/Hz}$ the Eddington flux 
of our final model (\Teffw{45\,000}, \loggw{4.46}, cf., Table\,\ref{tab:para}) at $\lambda 5454\,\mathrm{\AA}$, and 
$m_\mathrm{v_0} = m_\mathrm{V} - 2.175  \times c$ the extinction corrected apparent magnitude. 
Using $E_\mathrm{B-V} = 0.038 \pm 0.010$, (Sect.\,\ref{sect:obs}, Fig.\,\ref{fig:ebv}) and
$c = 1.47 E_\mathrm{B-V}$ from the standard stellar extinction curve \citep[e.g.,][]{osterbrock1974},
we calculate $m_\mathrm{v_0} = 8.17 \pm 0.03$.
This yields a distance of $d = 496^{+82}_{-83}$\,pc, in good agreement with previously determined values
(Table\,\ref{tab:para}), and a height below the Galactic plane of $H = 163 \pm 27\,\mathrm{pc}$.

Using the \textit{Gaia} distance
and 
$m_\mathrm{V} = 8.287 \pm 0.0024$ \citep{landoltuomoto2007}, the absolute magnitude is 
$M_\mathrm{V} = -0.30^{+0.05}_{-0.06}$. 
Using the bolometric-correction formula of \citet{martinsetal2005}, we get 
$BC = 27.58 - 6.8 \times \log{T_\mathrm{eff}} = -4.06^{+0.07}_{-0.06}$ , 
the luminosity can be approximated using the following formula taken from \citet{martinsetal2005},

\begin{equation}
        \log{\frac{L}{L_{\odot}}} = -0.4 \times \left( M_\mathrm{V} + \mathrm{BC} - M_{\odot}^{\mathrm{bol}} \right) = 3.64 \pm 0.05 \quad ,
\end{equation}

\noindent
where 
$M_{\odot}^{\mathrm{bol}} = 4.74$ 
is the solar bolometric luminosity \citep{torresetal2010} and 
$L_\odot = 3.828 \times 10^{33}$\,erg/s is
the solar luminosity \citep{iau2015b2}.

\noindent
The radius of \hd is derived using the Stefan-Boltzmann law

\begin{equation}
\frac{R}{R_\odot} = \left( \frac{T_\odot}{T} \right)^2 \left( \frac{L}{L_\odot}\right)^{\frac{1}{2}} = 1.09^{+0.11}_{-0.10} \quad .
\end{equation}

\noindent
This radius is only slightly larger than the value by \citet[][$1.05\pm 0.06\,R_\odot$]{krtickaetal2019}, 
and is in agreement with the value calculated by \citet[][$1.45\pm 0.25\,R_\odot$]{kudritzkisimon1978}.
Finally, the mass of \hd is calculated using

\begin{equation}
        \frac{M}{M_\odot} = \left(\frac{g}{g_\odot}\right)\left(\frac{R}{R_\odot}\right)^2 = 1.14^{+0.30}_{-0.24}  \quad .
\end{equation}

\noindent
This mass is lower than the previously determined values of 
\citet[][$1.46\pm 0.32\,M_\odot$]{krtickaetal2019} and 
\citet[][$1.75\pm 1.00\,M_\odot$]{kudritzkisimon1978}
but agrees within the error limits.

\section{Results and conclusions}
\label{sect:results}

We performed a detailed spectral analysis of \hd by means of advanced NLTE model-atmosphere techniques
based on the available ORFEUS (TUES), HST/GHRS, and UVES spectra (Table\,\ref{tab:spectra}).
We found \Teffw{45\,000\pm 1\,000} and \loggw{4.46\pm 0.10}, which is within the error limits (Table\,\ref{tab:parameters})
and in good agreement with the results of \citet{krtickaetal2019} (Table\,\ref{tab:parameters}).
Our H/He mass-fraction ratio of 0.20/0.78 (Fig.\,\ref{fig:HHe_abund}) agrees only marginally with 
the ratio of 0.25/0.74 (Table\,\ref{tab:parameters}) that was found by \citet{krtickaetal2019}.

\begin{table}
\begin{center}
\caption{Parameters of \hd.}
\label{tab:parameters}
\setlength{\tabcolsep}{.25em}
\begin{tabular}{lr@{$\pm$}lr@{$\pm$}lcr@{$\pm$}lr@{$\pm$}lr@{.}l}                
\hline
\hline
\noalign{\smallskip}
 & 
\multicolumn{3}{c}{this work} & 
\multicolumn{7}{c}{\citet{krtickaetal2019}} \\
\hline
\noalign{\smallskip}
\Teff\,/\,K                    & \multicolumn{3}{c}{$45\,000 \pm 1\,000$}                & \multicolumn{7}{c}{$45\,900 \pm 800$}              \\
$\log (g\,/\,\mathrm{cm/s^2})$ & \multicolumn{3}{c}{\hspace{0.45em}$4.46 \pm 0.10$}      & \multicolumn{7}{c}{\hspace{0.95em}$4.56 \pm 0.08$} \\
\noalign{\smallskip}
$L$\,/\,$L_\odot$               & \multicolumn{3}{c}{\hspace{0.69em}$4360^{+509}_{-456}$}  & \multicolumn{7}{c}{~}                              \\
\noalign{\smallskip}
$M$\,/\,$M_\odot$               & \multicolumn{3}{c}{\hspace{0.69em}$1.14^{+0.30}_{-0.24}$} & \multicolumn{7}{c}{\hspace{0.85em}$1.46 \pm 0.32$} \\
\noalign{\smallskip}
$R$\,/\,$R_\odot$               & \multicolumn{3}{c}{\hspace{0.69em}$1.09^{+0.11}_{-0.10}$} & \multicolumn{7}{c}{\hspace{0.95em}$1.05 \pm 0.06$} \\
\noalign{\smallskip}
$d$\tablefootmark{a}\,/\,pc    & \multicolumn{3}{c}{\hspace{0.69em}$496^{+0.82}_{-0.83}$}  & \multicolumn{7}{c}{\hspace{0.2em}$508 \pm 17$}     \\
\noalign{\smallskip}
$v_\mathrm{rot}\,/\,\mathrm{km/s}$ & \multicolumn{3}{c}{ 37.5  $\pm$ 5.0} & \multicolumn{7}{c}{$\hspace{0.4em}$40 $\pm$ 5}\\
\hline  
\end{tabular}
\begin{tabular}{lr@{$\pm$}lr@{$\pm$}lcr@{$\pm$}lr@{$\pm$}l}                
& \multicolumn{8}{c}{Abundances} \\
\cline{2-10}
& \multicolumn{4}{c}{this work} & 
\multicolumn{5}{c}{\citet{krtickaetal2019}} \\
\cline{2-10}
\multicolumn{9}{c}{~} \vspace{-5mm}\\
 \vspace{-2mm}\\
&
\multicolumn{2}{c}{num} & 
\multicolumn{2}{c}{mass} &
\multicolumn{3}{c}{num} &
\multicolumn{2}{c}{mass} \\
\hline
H  & $-$0.30 & 0.03                & $-$0.70 & 0.03                 && \multicolumn{2}{l}{$\hspace{3.1mm}-$0.24} & \multicolumn{2}{l}{$\hspace{3.1mm}-$0.60} \\
He & $-$0.31 & 0.03                & $-$0.11 & 0.03                 &&  $\hspace{3.1mm}-$0.37 & 0.04             &  $\hspace{3.1mm}-$0.13 & 0.02             \\
C  & $-$4.92 & 0.1                 & $-$4.25 & 0.1                  && \multicolumn{2}{l}{$-$4.44\tablefootmark{b}}               & \multicolumn{2}{l}{$-$3.72\tablefootmark{b}}  \\
N  & $-$3.10 & 0.1                 & $-$2.36 & 0.1                  &&  $\hspace{3.1mm}-$3.34 & 0.2              &  $\hspace{3.1mm}-$2.56 & 0.2             \\
O  & $-$4.90 & 0.1                 & $-$4.10 & 0.1                  &&  $\hspace{3.1mm}-$4.84 & 0.2              &  $\hspace{3.1mm}-$4.00 & 0.2             \\
Ne & $-$3.26 & 0.1                 & $-$2.36 & 0.1                  && \multicolumn{2}{c}{--}                    & \multicolumn{2}{c}{--}                   \\
Mg & $-$4.24 & 0.1                 & $-$3.26 & 0.1                  && \multicolumn{2}{l}{$\hspace{3.1mm}-$4.44} & \multicolumn{2}{l}{$\hspace{3.1mm}-$3.42}\\
Al & $-$5.21 & 0.1                 & $-$4.18 & 0.1                  &&  $\hspace{3.1mm}-$5.14 & 0.4              &  $\hspace{3.1mm}-$4.07 & 0.4             \\
Si & $-$3.72 & 0.1                 & $-$2.68 & 0.1                  &&  $\hspace{3.1mm}-$4.34 & 0.4              &  $\hspace{3.1mm}-$3.25 & 0.4             \\
P  & $-$5.58 & 0.1                 & $-$4.50 & 0.1                  && \multicolumn{2}{c}{--}                    & \multicolumn{2}{c}{--}                   \\
S  & $-$3.61 & 0.1                 & $-$2.51 & 0.1                  && \multicolumn{2}{c}{--}                    & \multicolumn{2}{c}{--}                   \\
Cr & $-$5.42 & 0.2                 & $-$4.11 & 0.2                  && \multicolumn{2}{c}{--}                    & \multicolumn{2}{c}{--}                   \\
Mn & $-$5.52 & 0.2                 & $-$4.18 & 0.2                  && \multicolumn{2}{c}{--}                    & \multicolumn{2}{c}{--}                   \\
Fe & $-$3.57 & 0.1                 & $-$2.23 & 0.1                  &&  $\hspace{3.1mm}-$4.14 & 0.2              &  $\hspace{3.1mm}-$2.75 & 0.2             \\
Ni & $-$4.58 & 0.1                 & $-$3.22 & 0.1                  &&  $\hspace{3.1mm}-$5.24 & 0.1              &  $\hspace{3.1mm}-$3.83 & 0.1             \\
\hline
\end{tabular}
\tablefoot{Our values are compared to the analysis of \citet{krtickaetal2019}. 
``<'' denotes an upper limit, ``--'' indicates that no abundance was determined.
The abundances are given as log number fraction (num) and log mass fraction (mass). 
\tablefoottext{a}\mbox{Spectroscopic distance}.
\tablefoottext{b}\mbox{Upper limit}.}
\end{center}
\end{table}

\begin{figure}
   \resizebox{\hsize}{!}{\includegraphics{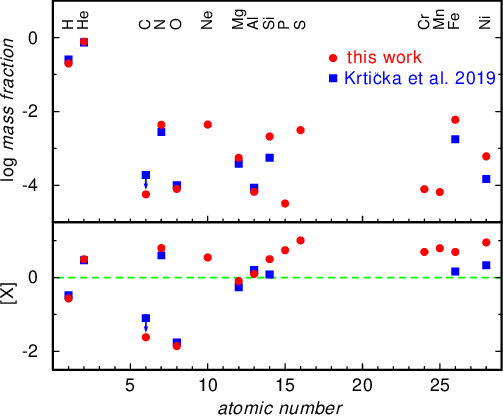}}
           \caption
           {Comparison of photospheric abundances of \hd with the results of \citet{krtickaetal2019}. 
            The uncertainties of the abundances are given in Table\,\ref{tab:parameters}. 
            Arrows indicate upper limits.
            \emph{Top}: Abundances given as logarithmic mass fraction. 
            \emph{Bottom}: Abundance ratios to the respective solar values \citep{asplundetal2009,scottetal2015a,scottetal2015b}.
                           [X] denotes the log (mass fraction\,/\,solar mass fraction) of species X.
                           The dashed green line indicates solar abundances.
            }
   \label{fig:abund}
\end{figure}

We determined the photospheric abundances of H, He, C, N, O, Me, Al, Si, P, S, Cr, Mn, Fe, and Ni (Table\,\ref{tab:parameters},
Fig.\,\ref{fig:abund}). 
A comparison of the ORFEUS, HST/GHRS, and UVES observations with
our final TMAP spectrum is shown in Figs.\,\ref{fig:orfeusall} - \ref{fig:uvesall}.

\begin{figure}
\centering
\includegraphics[scale=1]{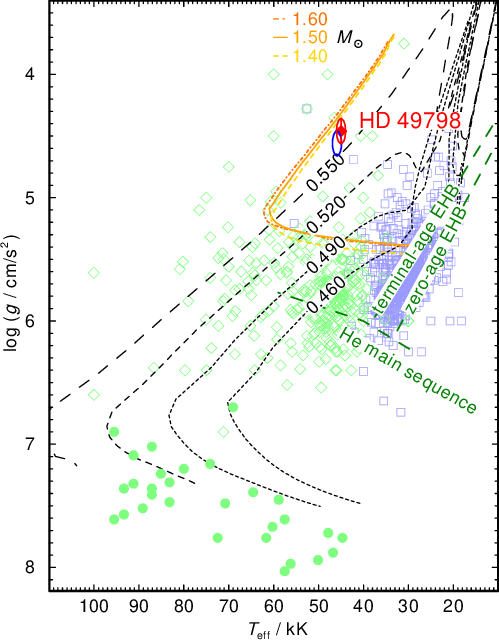}
\caption{Location of \hd (red error ellipse) in the log\,\Teff\!-\,\logg\ plane compared 
with three MESA models for the evolution of He cores with different masses 
\citep[orange tracks, evolution towards lower \logg, taken from][]{brooksetal2017}.
The blue error ellipse shows the error limits of \citet{krtickaetal2019}.
Positions of 
sdB \citep[blue, open squares]{geieretal2017}, 
sdO stars \citep[green, open rhombs]{geier2020}, 
and H-rich WDs \citep[green, dashed lines]{gianninasetal2010} are indicated. 
Four post-EHB evolutionary tracks, labeled with the stellar mass in $M_\odot$, 
\citep[blue, dashed lines for $Y = 0.459 \approx 1.6 \times Y_\odot$]{dormanetal1993} are shown
to indicate their evolution. 
The dashed green lines indicate the He main sequence and the zero- and terminal-age EHB.}
\label{fig:evolution}
\end{figure}

\hd remains interesting regarding its origin and
future evolution as it seems to deviate from most observations made
regarding hot subdwarfs. Its low surface gravity of \loggw{4.46} in
combination with the high \Teffw{45\,000} places it
outside the known region in the log\,\Teff\!-\,\logg\ plane
inhabited by sdOs (Fig.\,\ref{fig:evolution}).
Thus, \hd did not pass the classical
horizontal-branch (HB) or extended horizontal-branch (EHB) evolutionary scenario.

\hd is a stripped, intermediate-mass He star. 
\citet{brooksetal2017} calculated three MESA\footnote{Modules for Experiments in Stellar Astrophysics} sdO models
\citep{paxtonetal2011, paxtonetal2013, paxtonetal2015}
for a ZAMS star with $M_\mathrm{ZAMS} = 7.15\,M_\odot$ and different mass-loss rates to match
\Teffw{47\,500 \pm 2000} and \loggw{4.25 \pm 0.2} of \citet{kudritzkisimon1978}. Their
evolutionary tracks (digitized with Dexter\footnote{\url{http://dc.zah.uni-heidelberg.de/sdexter}, \url{http://soft.g-vo.org/dexter}}) 
are shown in Fig.\,\ref{fig:evolution}. Our values of \Teffw{45\,000 \pm 1000} and \loggw{4.46 \pm 0.1} show \hd just below these
tracks, i.e., at a He-core mass of $\sla 1.4\,M_\odot$. 

The CNO abundances of \hd are in line with yields of the CNO cycle and 3\,$\alpha$ nucleosynthesis. 
Significantly over-solar abundances of heavy elements like Cr, Mn, Fe, and Ni
may be the result of diffusion. For \hd, \citet{krtickaetal2019} measured a mass-loss rate of 
$\dot{M} = 2.1 \times 10^{-9} M_\odot/\mathrm{yr}$, which is too high for an efficient diffusion.
This mass-loss rate is slightly lower than the value for an assumed 
solar composition $2.7 \times 10^{-9} M_\odot/\mathrm{yr}$ \citep[cf.,][]{krtickaetal2019} and, thus,
subsolar heavy-metal abundances would be expected.

However, we searched for lines of trans-iron elements \citep[cf.,][]{hoyeretal2017,hoyeretal2018,loeblingetal2020,rauchetal2020},
but this was entirely unsuccessful. This was expected at the high $v_\infty = 1\,570\,\mathrm{km/s}$ \citep{krtickaetal2019}
and a height of the photosphere of $\approx 5\times 10^4\,\mathrm{km}$.

The compact companion of \hd has a rotational period of 13.2\,s \citep{israeletal1997}
and a very high spin-up rate.
\citet{mereghettietal2021} analyzed results of \textit{XMM-Newton}\footnote{X-ray Multi Mirror Mission} pointings 
and found a constant spin-up rate of $\dot{P} = (-2.17 \pm 0.01) \times 10^\mathrm{-15}\,\mathrm{ss}^{-1}$ , 
which led them to presume it was a WD covered with helium-rich material accreted from \hd. 
Recently, \citet{rigosellietal2023} measured $\dot{P} = (-2.28 \pm 0.02) \times 10^\mathrm{-15}\,\mathrm{ss}^{-1}$
from NICER\footnote{Neutron Star Interior Composition Explorer}, \textit{XMM-Newton}, and
ROSAT\footnote{R\"ontgen Satellit} observations (together over a period of $\approx 30$ years).
Such a high spin-up cannot be explained by accretion alone due to the low mass transfer in this
binary \citep{mereghettietal2016}. The scenario proposed by \citet{popovetal2018} is more likely,
in which the WD, which is only a few million years old, is contracting.
An ongoing analysis of photometric TESS\footnote{Telescope Encoder and Sky Sensor} data
(Schaffenroth et al\@. in prep.) will further improve the orbital parameters and our understanding
of \hd.

We discovered a very good spectrum of \hd (Fig.\,\ref{fig:orfeusall}) in the ORFEUS database
(\url{https://archive.stsci.edu/tues/search.php}) -- observed but never analyzed or published.
Without this observation, \citet{krtickaetal2019} could not evaluate 
\ionww{C}{iii}{1174.93 - 1176.37} and \ionww{C}{iv}{1168.89, 1168.98}
and derived only an upper C-abundance limit from the UVES observation. Their upper limit, however, is close to
the value that we determined (Fig.\,\ref{fig:abund}).
At least the prominent \ion{N}{iv} and \ion{S}{vi} P\,Cygni profiles (Fig. \ref{fig:orfeusall}) deserve further analysis;
this is not possible with our hydrostatic TMAP.
However, we expect there to be many other ``forgotten'' spectra in other archives, ready to be analyzed.

\begin{acknowledgements}
We thank 
Klaus Werner for helpful comments, discussions, and for careful reading of the manuscript;
Jiri Krti{\v{c}}ka for putting the reduced optical UVES spectrum, that had been used in
\citet{krtickaetal2019}, at our disposal;
and Klaas de Boer, for many long-time-ago discussions about UV spectroscopy (especially about ORFEUS).

This research has made use of the SIMBAD database, operated at CDS, Strasbourg, France,
and of NASA's Astrophysics Data System.

The TIRO (\url{http://astro.uni-tuebingen.de/~TIRO}) tool and the 
    TMAD (\url{http://astro.uni-tuebingen.de/~TMAD}) service
were constructed as part of the T\"ubingen project 
( \url{https://uni-tuebingen.de/de/122430})
of the German Astrophysical Virtual Observatory
(GAVO, \url{http://www.g-vo.org}).

This work used the profile-fitting procedure OWENS developed by M\@. Lemoine and the FUSE French Team.
\end{acknowledgements}

\bibliographystyle{aa}
\bibliography{hd49798}

\addtocounter{section}{1}
\appendix
\section{Additional figures}
\label{appendixB}
\clearpage

\begin{figure*}[h!]
 \centering
 \includegraphics{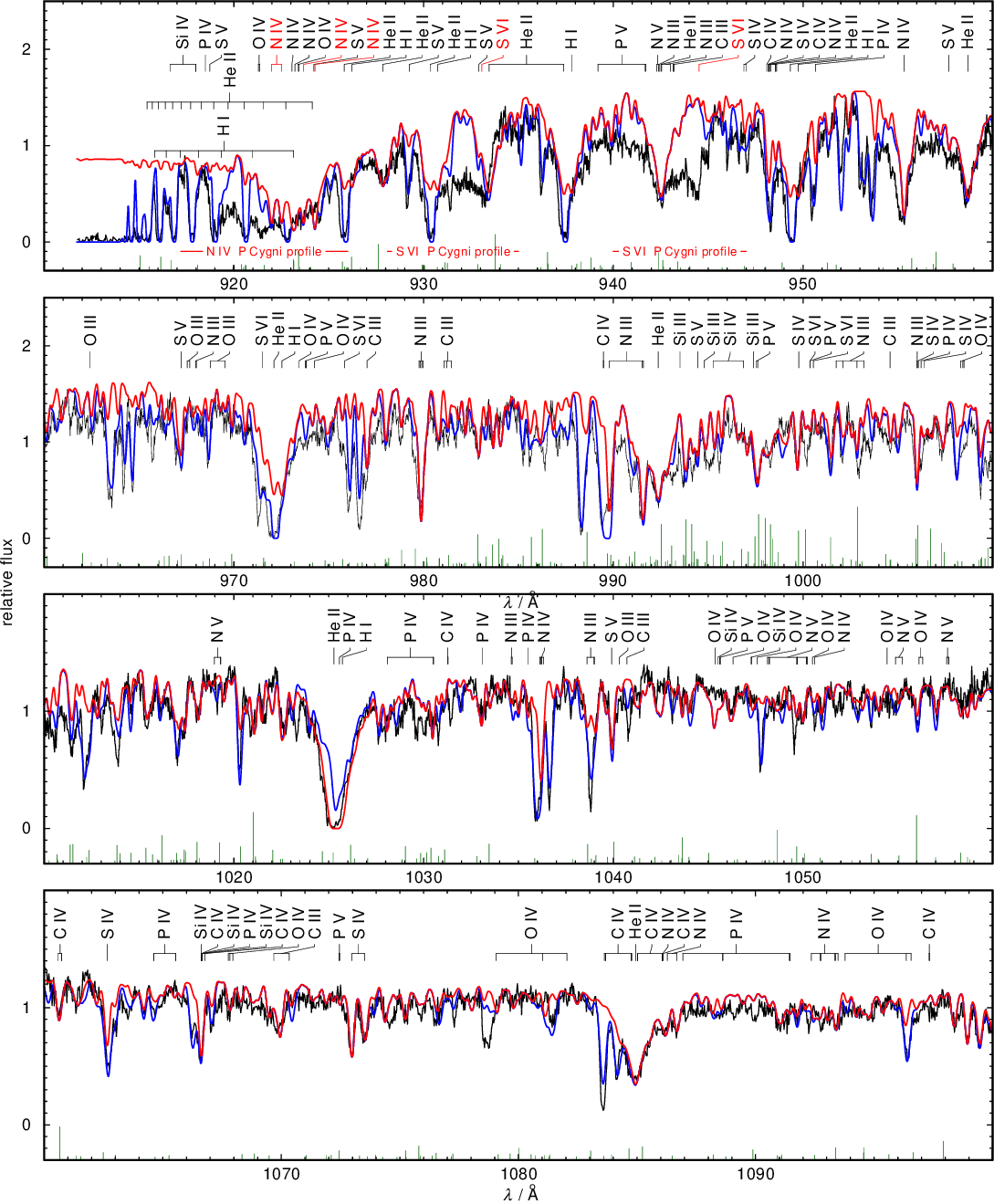} 
   \caption[]{ORFEUS\,II observation (black) compared with our best model with (blue) and without (red)
              interstellar line absorption considered. 
              $v_\mathrm{rad} = 140\,\mathrm{km/s}$ and
              $v_\mathrm{rot} = 37.5\,\mathrm{km/s}$ are applied to our model.
              Stellar lines are identified at top. 
              Prominent P\,Cyg profiles of 
              \Jonww{N}{iv}{922.0, 922.5, 923.1, 924.3}, 
              \Jonww{N}{v}{1238.8, 1242.8}, and 
              \Jonww{S}{vi}{933.4, 944.5} are indicated by horizontal bars.
              The green marks at the bottom of each panel 
              indicate wavelengths and strengths (mark lengths $\sim \log gf$ value of the respective line) 
              of prominent iron-group (Ca-Ni) lines.
              } 
   \label{fig:orfeusall}
\end{figure*}

\begin{figure*}
\centering
\ContinuedFloat
 \includegraphics{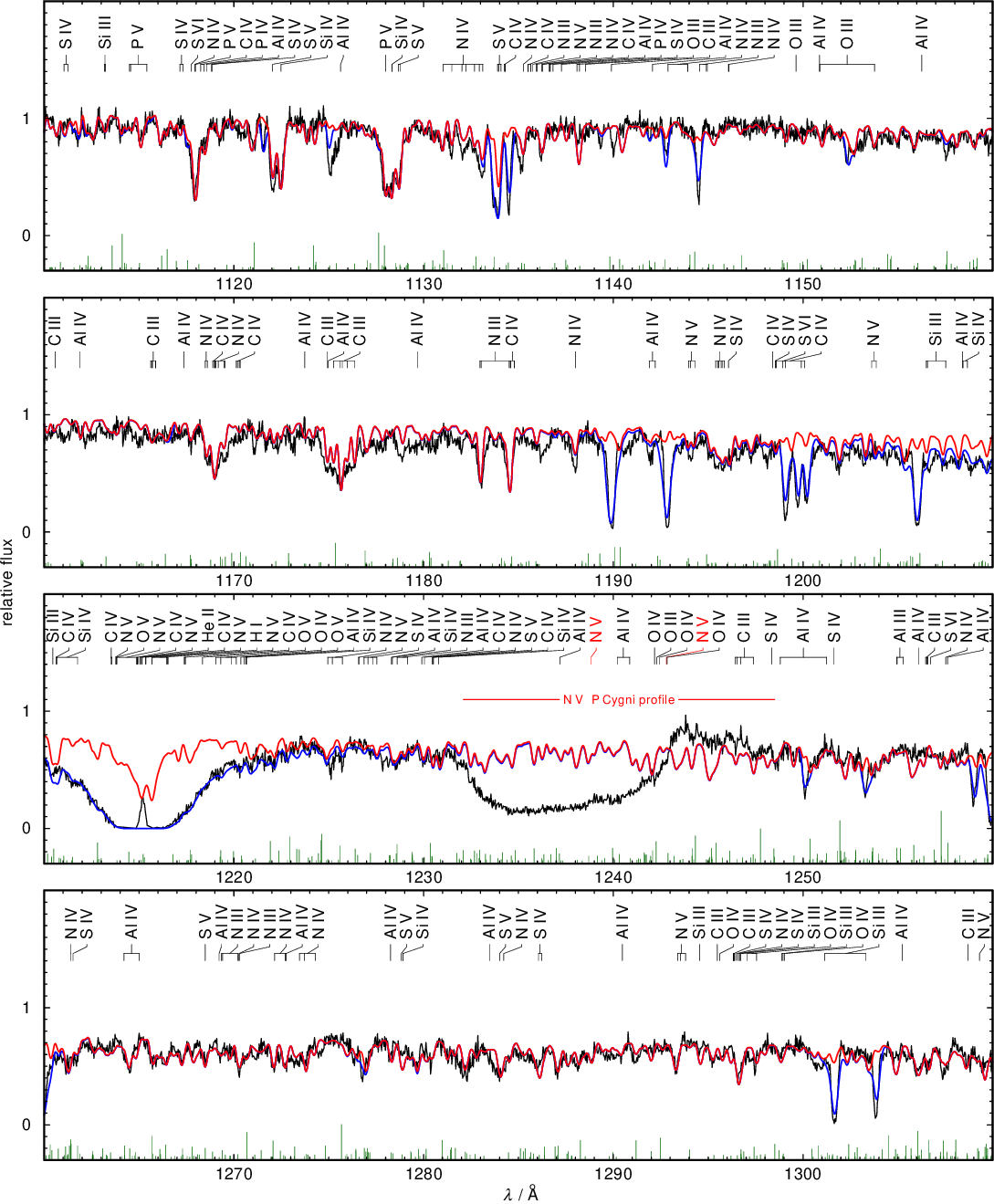} 
   \caption[]{Continued.
             } 
\end{figure*}

\begin{figure*}
\centering
\ContinuedFloat
 \includegraphics{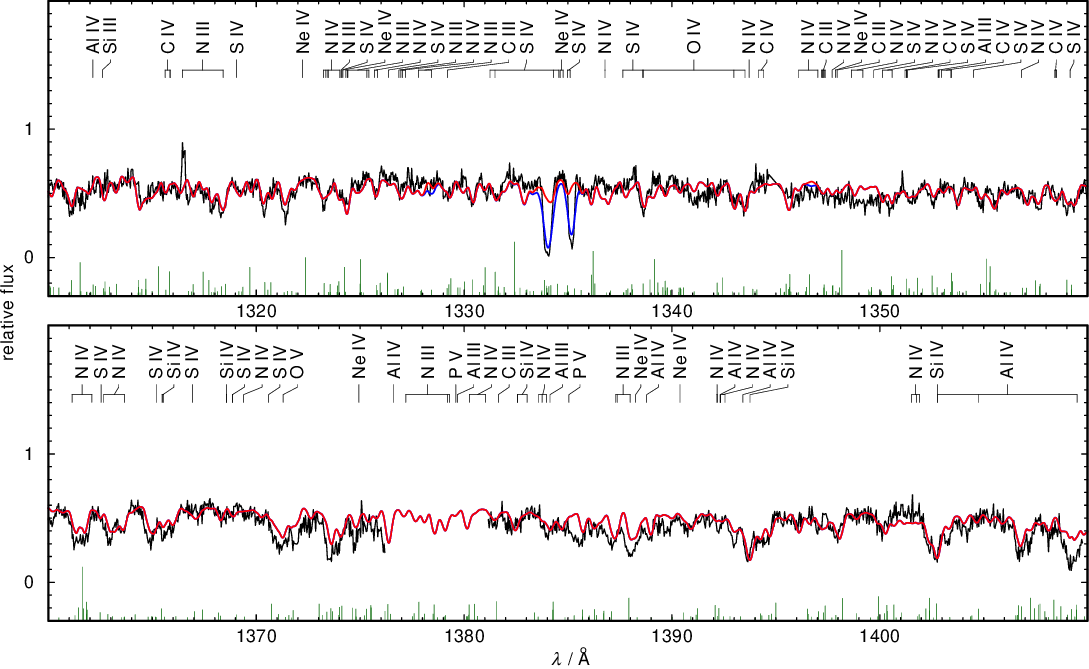}
 \vspace*{-6mm}
   \caption[]{Continued.
             } 
\end{figure*}

\begin{figure*}[h!]
 \centering
 \includegraphics{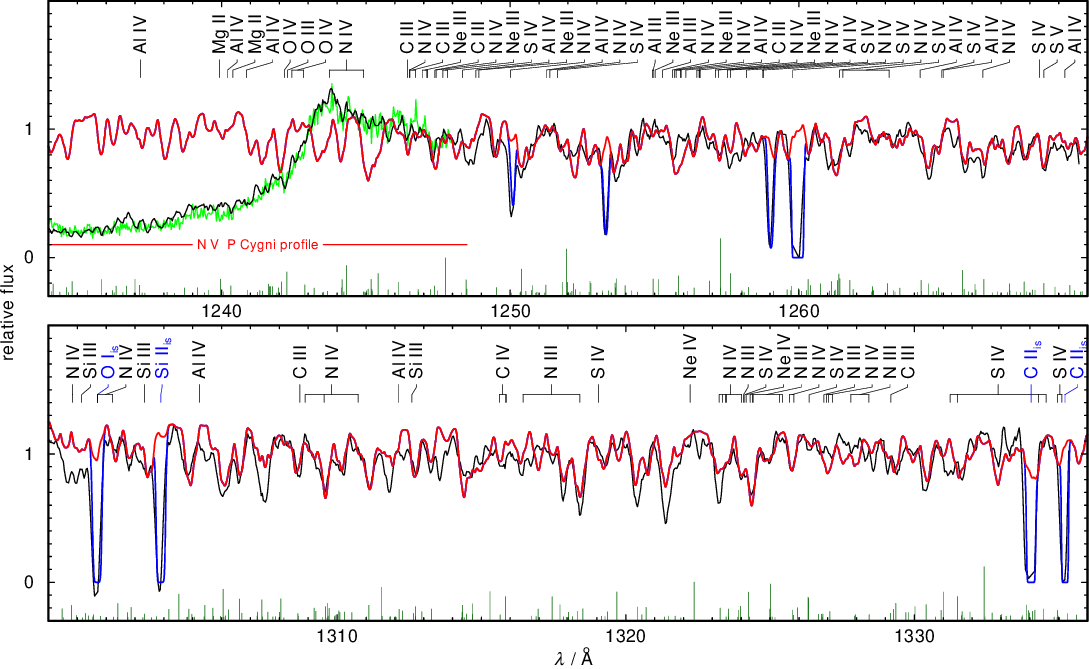} 
 \vspace*{-6mm}
   \caption[]{Like Fig.\,\ref{fig:orfeusall} but for the GHRS observations.
              In the range of the \Jonww{N}{v}{1238.8, 1242.8} P\,Cygni profile,
              the ORFEUS\,II spectrum  (green) is shown for comparison.
              } 
   \label{fig:ghrsall}
\end{figure*}

\begin{figure*}
\centering
\ContinuedFloat
 \includegraphics{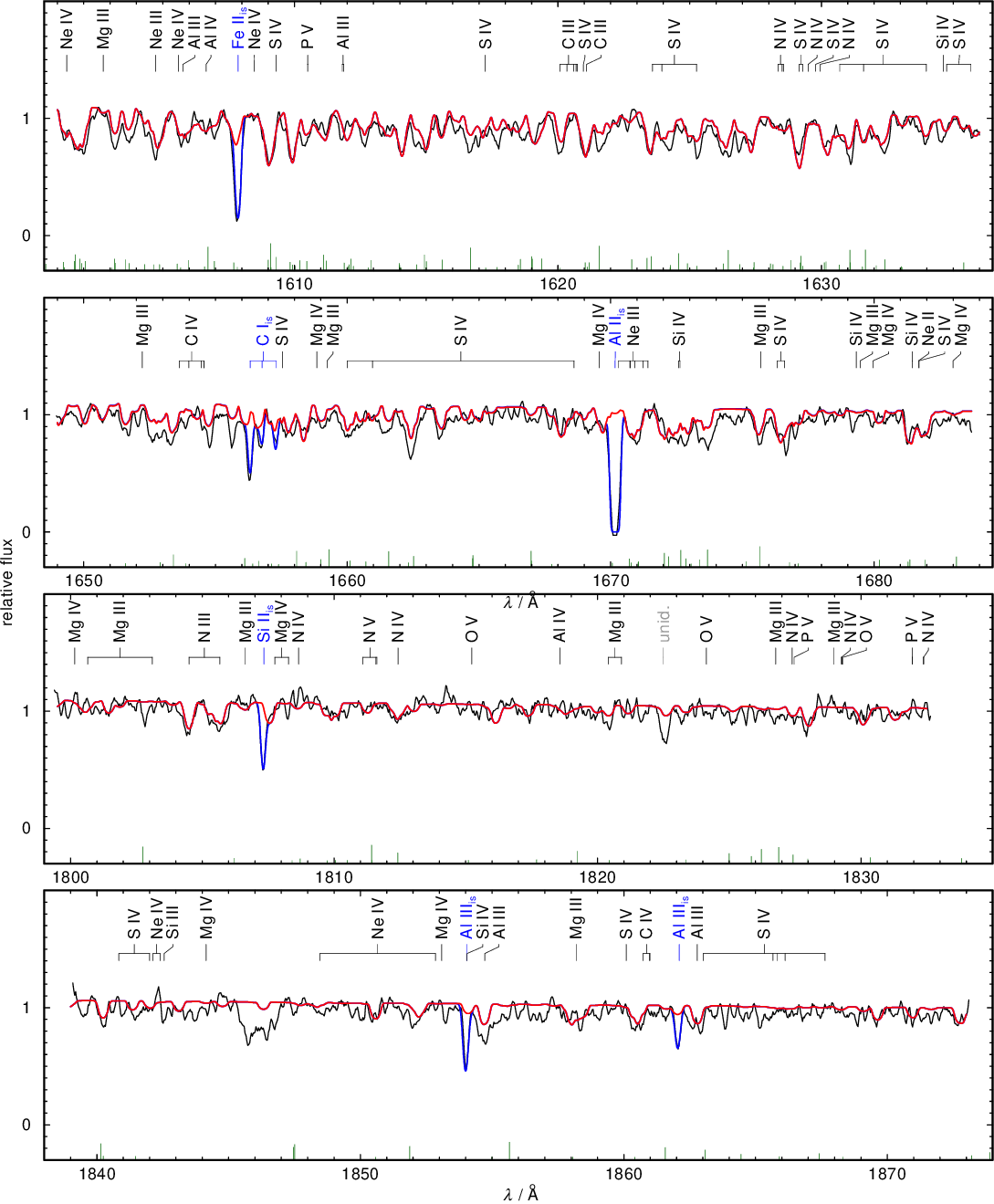} 
   \caption[]{Continued.
             } 
\end{figure*}

\begin{figure*}
 \centering
 \includegraphics{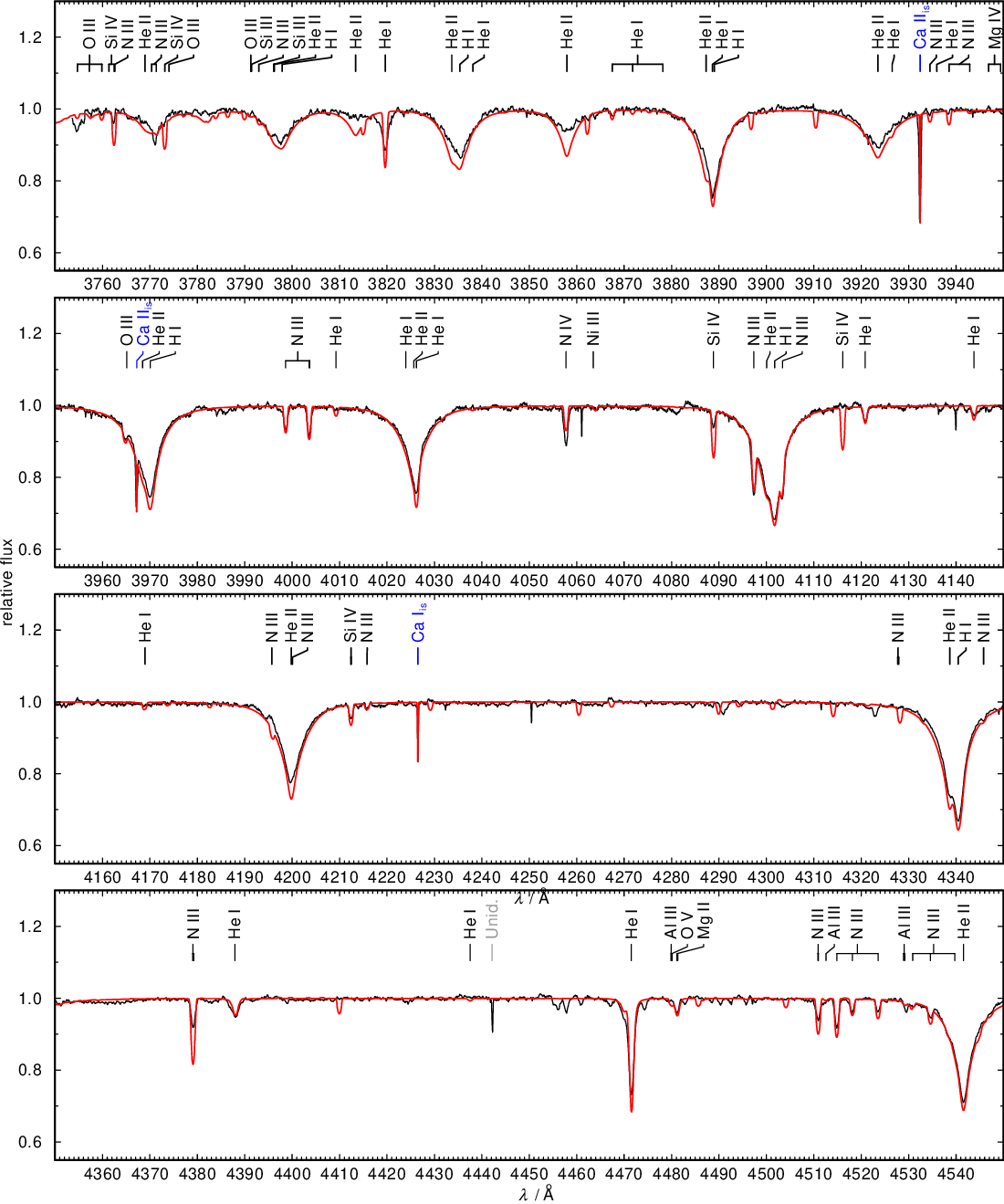} 
   \caption[]{Like Fig.\,\ref{fig:orfeusall} but for the UVES observation.
              } 
   \label{fig:uvesall}
\end{figure*}

\begin{figure*}
\centering
\ContinuedFloat
 \includegraphics{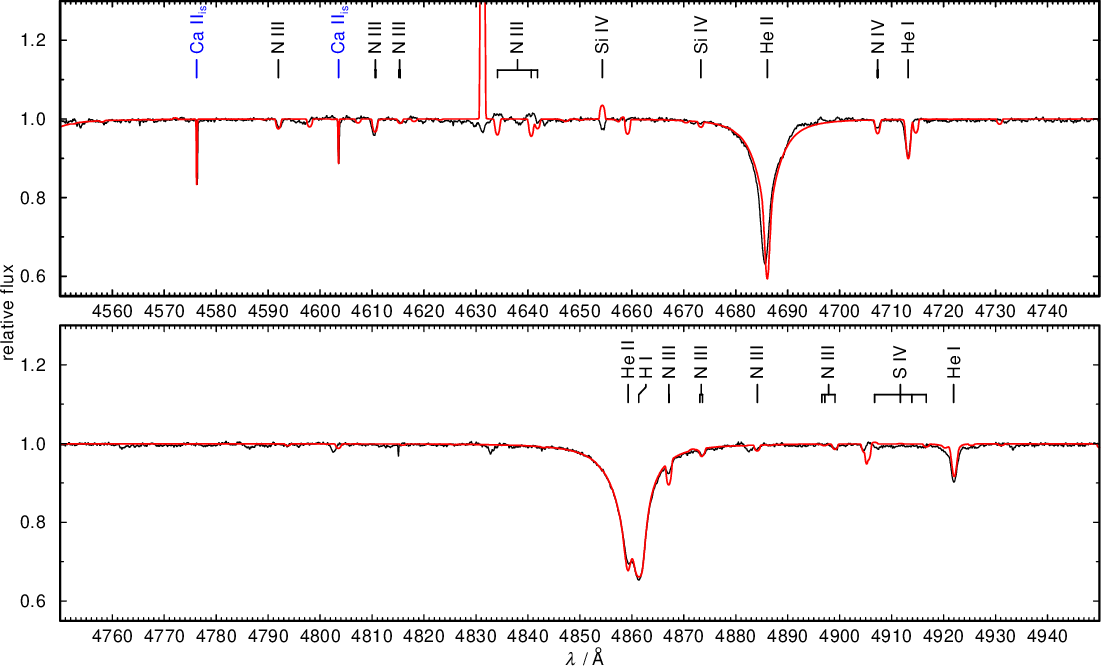} 
   \caption[]{Continued.
             } 
\end{figure*}

\end{document}